\documentclass[fleqn,usenatbib]{mnras}

\usepackage{newtxtext,newtxmath}
\usepackage[T1]{fontenc}

\def \rxte{{\it RXTE}}
\def \inte {{\it INTEGRAL}}
\def \swift {{\it Swift}}

\def \xmm {{\it XMM}-Newton}
\def \src{{RX J0440.9+4431}}

\def \swiftbat{{\it Swift}/BAT}
\def \nustar{{\it NuSTAR}}

\def \nicer{{\it NICER}}
\def \rosat{{\it ROSAT}}

\def \fermigbm{{\it Fermi}/GBM}

\def \astrosat{\textit{AstroSat}}
\def \ixpe{{\it IXPE}}
\def \insighthxmt{{\it Insight}/HXMT}

\newcommand{\psec}{s$^{-1}$}
\newcommand{\erg}{erg cm$^{-2}$ s$^{-1}$} % unit of Flux
\newcommand{\lum}{erg s$^{-1}$}

\DeclareRobustCommand{\VAN}[3]{#2}
\let\VANthebibliography\thebibliography
\def\thebibliography{\DeclareRobustCommand{\VAN}[3]{##3}\VANthebibliography}

\usepackage{graphicx}	% Including figure files
\usepackage{amsmath, bm}	% Advanced maths commands
\usepackage{ulem}	
\usepackage{booktabs}
\title[AstroSat observations of \src]{Probing the energy and luminosity-dependent spectro-timing properties of \src{} with \astrosat}

\author[Sharma et al.]{
Rahul Sharma,$^{1}$\thanks{E-mail: rahul1607kumar@gmail.com}
Manoj Mandal,$^{2}$\thanks{E-mail: manojmandal@mcconline.org.in}
Sabyasachi Pal,$^{2}$
Biswajit Paul,$^{1}$
G. K. Jaisawal$^{3}$
and Ajay Ratheesh$^{4}$
\\
$^{1}$Raman Research Institute, C.V. Raman Avenue, Sadashivanagar, Bengaluru, Karnataka 560080, India\\
$^{2}$Midnapore City College, Bhadutala, West Bengal, India 721129\\
$^{3}$DTU Space, Technical University of Denmark, Elektrovej 327-328, DK-2800 Lyngby, Denmark\\
$^{4}$INAF Istituto di Astrofisica e Planetologia Spaziali, Via del Fosso del Cavaliere 100, 00133 Roma, Italy
}

\date{Accepted XXX. Received YYY; in original form ZZZ}

\pubyear{2024}

\begin{document}
\label{firstpage}
\pagerange{\pageref{firstpage}--\pageref{lastpage}}
\maketitle

\begin{abstract}
The Be/X-ray binary pulsar RX J0440.9+4431 went through a giant outburst in December 2022 with a peak flux of $\sim$2.3 Crab in 15--50 keV. We studied the broad-band timing and spectral properties of \src{} using four \astrosat{} observations, where the source transited between subcritical and supercritical accretion regimes. Pulsations were detected significantly above 100 keV. The pulse profiles were found to be highly luminosity- and energy-dependent. A significant evolution in the pulse profile shape near the peak of the outburst indicates a possible change in the accretion mode and beaming patterns of \src. The rms pulsed fraction was luminosity- and energy-dependent, with a concave-like feature around 20--30 keV. The depth of this feature varied with luminosity, indicating changes in the accretion column height and proportion of reflected photons. The broad-band continuum spectra were best fitted with a two-component Comptonization model with a blackbody component or a two-blackbody component model with a thermal Comptonization component. A quasi-periodic oscillation at 60 mHz was detected at a luminosity of $2.6 \times 10^{37}$ \lum, which evolved into 42 mHz at $1.5 \times 10^{37}$ \lum. The QPO rms were found to be energy dependent with an overall increasing trend with energy. For the first time, we found the QPO frequency varying with photon energy in an X-ray pulsar, which poses a challenge in explaining the QPO with current models such as the Keplarian and beat frequency model. Hence, more physically motivated models are required to understand the physical mechanism behind the mHz QPOs. 

\end{abstract}


%%%%%%%%%%%%%%%%%%%%%%%%%%%%%%%%%%%%%%%%%%%%%%%%%%
%%%%%%%%%%%%%%%%% BODY OF PAPER %%%%%%%%%%%%%%%%%%

\section{Introduction}
\label{intro}

One significant subgroup of high-mass X-ray binaries is Be/X-ray binaries \citep[BeXRBs;][]{Liu06}. These systems consist of a neutron star in an eccentric orbit around a Be star \citep{Re11, Mushtukov2022}. Classical Be stars are B-type stars with emission lines in their spectra (where the letter ``e'' indicates emission), especially hydrogen lines from the Balmer series, and sometimes helium and iron lines (\citealt{Porter2003, Re05}). The optical emission primarily comes from the Be star, while the X-ray emission reflects the physical state near the neutron star.  
Be/X-ray binaries are recognized for featuring two distinct types of X-ray outbursts: Type-I and Type-II. Type-I X-ray outbursts typically originate around the periastron point of a neutron star, whereas Type-II X-ray outbursts, which may be related to the warping of the outermost region of the circumstellar disc, can take place at any orbital phase \citep{Oka13}.

The resonant scattering of continuum photons with electrons produces absorption-like features in the energy spectrum. In the presence of strong magnetic fields, electron orbital energy is quantized into Landau levels \citep{Meszaros1992}. In high-mass X-ray binaries (HMXBs), cyclotron resonant scattering features (CRSFs) are mostly observed in the energy range of 10--100 keV \citep{Staubert2019}. The cyclotron line energy ($E_{c}$) is used to estimate the magnetic field strength of a neutron star, following the relation $E_{c} = 11.6 B_{12} n (1+z)^{-1}$ keV, where $B_{12}$ is the magnetic field strength in the unit of 10$^{12}$ G, $n=1,2,3,...$ is the number of Landau levels, and $z$ is the gravitational redshift. Variations in the geometry and dynamics of accretion flow across the magnetic poles of the neutron star can affect these cyclotron line scattering features. 

Strong outbursts, often recorded from some X-ray pulsars \citep[e.g., 1A 0535+262, EXO 2030+375, 4U 0115+63, KS 1947+300;][]{Reig2013, Mandal2022}) can produce such high X-ray luminosities that the accretion column begins to rise due to emerging radiation, halting the accretion above the neutron star surface via a radiation-dominated shock. This threshold luminosity is known as the `critical luminosity' \citep{Mushtukov2015}. The critical luminosity plays a crucial role in defining two distinct accretion regimes, above which the beaming patterns and accretion mechanism of the pulsars evolve significantly. Below the critical luminosity, accretion occurs in the subcritical regime, where a ``pencil beam'' of X-ray emission forms as accreting material falls freely onto the neutron star's surface \citep{Basko1975}. In this case, photons propagate along the magnetic field lines as emission escapes from the top of the column in the pencil beam pattern \citep{Burnard1991}. At higher luminosities ($L_x\ge10^{37}$ erg s$^{-1}$), an accretion-dominated shock is thought to appear close to the critical luminosity, leading to a change in beaming patterns and a spectral shape. In the supercritical regime, the radiation pressure becomes strong enough to stop the accreting matter above the neutron star, resulting in a radiation-dominated shock \citep{Basko1976, Becker2012}. This regime is characterized by fan-beam patterns or a combination of pencil and fan-beam patterns. However, the actual beaming patterns can sometimes be more complex than the simple fan or pencil beam patterns \citep{Kraus1995, Blum2000, Becker2012}.

\subsection{\src}

The Be/X-ray binary pulsar \src{} was discovered with \rosat{} during a galactic plane survey \citep{Motch1997}. The long-term optical/infrared study revealed that the optical counterpart of the source is BSD 24-491/LS V +44 17, with a spectral type of B0.2V \citep{Motch1997, Re05}. The distance to the source was initially found to be 3.3$\pm$0.5 kpc \citep{Re05}, and later, a more precise distance of $\sim$2.44 kpc was suggested by \citet{Ba21} using data from {\it Gaia} DR3. From the spectroscopy results, \citet{Re05} reported the H$_{\alpha}$ line exhibiting a variable double peak shape, with a correlation observed between the infrared magnitude and equivalent width of the H$_{\alpha}$ line. The \rxte{}/PCA light curves confirmed an X-ray pulsation of 202.5$\pm$0.5 s in the 3--20 keV energy band \citep{Re99}. 

X-ray outbursts were detected from the source between 2010 and 2011. \citet{Ferrigno2013} estimated the orbital period of \src{} to be $\sim$150 days based on three consecutive X-ray outbursts in the \swift{}/BAT light curve. During the 2010 outburst, \cite{Ts12} used \rxte{}, \swift{}, and \inte{} to study several spectral and timing properties of \src{}. \cite{Ts12} reported a pulse period of 205.0$\pm0.1$ s during the 2010 September outburst. The pulse profile had a single peak feature and was nearly sinusoidal. The pulse profile was neither dependent on luminosity nor energy \citep{Ts12}. The \inte{} spectral analysis of the source detected a cyclotron line at 32 keV, estimating the magnetic field to be $3.2\times10^{12}$ G \citep{Ts12}. Based on the timing analysis of the \xmm{} data, \cite{Pa12} estimated the spin period to be $\sim$204.96 s, with an average spin-down rate of $6 \times 10^{-9}$ s s$^{-1}$ over 13 years. 

After more than a decade of inactivity, \src{} went through a giant outburst in December 2022, continuing for nearly four months. The X-ray flux reached nearly 2.3 Crab as observed by the \swiftbat{} (15--50 keV). The outburst was monitored in different wavelengths from radio to X-ray \citep{Nak22, Ma23, Pal23, Sa23, Kumari2023, Coley2023, vandenEijnden2024}. Using \inte{} and \nustar{} observations, \citet{Sa23} examined the spectra of \src{} and did not find any signature of CRSF. Two Comptonized components were used to characterize the source spectrum during the outburst \citep{Sa23}. The source also showed a significant evolution of pulse profile and spectral hardness near a critical luminosity, indicating a change in the accretion modes \citep{Mandal23, Li23}. 

Combined timing and spectral studies suggested a state transition from the subcritical to the supercritical accretion regime \citep{Mandal23, Sa23}, with changes in emission mechanisms and beaming patterns near the transition point. The variation of pulsed fraction (30--100 keV) with luminosity showed a change in correlation close to the critical luminosity, potentially related to the change in the accretion mechanism \citep{Li23}. As luminosity increased, the iron emission line evolved from a narrow to a broad feature, showing a substantial correlation with the X-ray \citep{Mandal23}. The magnetic field was also estimated to be of the order of 10$^{12}$ or 10$^{13}$ G based on the different theoretical models. A magnetic field of the order of 10$^{13}$ G was estimated from the torque luminosity model \citep{Sa23}. 

The X-ray polarization properties of \src\ were studied at two different luminosity levels during the outburst using Imaging X-ray polarimetry explorer (\ixpe{}) \citep{Doroshenko2023}. The source was likely observed in both supercritical and subcritical states, showing significantly different emission-region geometries associated with the presence of accretion columns and hot spots. The phase-resolved polarimetry indicated the presence of an unpulsed polarized component along with the polarized radiation from the pulsar, constraining the pulsar geometry with a pulsar inclination of $i_p \approx 108^\circ$ and a magnetic obliquity of $\theta_p \approx 48^\circ$ \citep{Doroshenko2023}. 

\citet{Malacaria24} reported the detection of $\sim$0.2 Hz quasi-periodic oscillation (QPO) in \src{} using data from \fermigbm, observed only above a certain luminosity and at specific pulse phases of the neutron star. This QPO is thought to originate when X-ray luminosity is near the supercritical accretion regime, where emission patterns change above a certain luminosity \citep{Malacaria24}. Similarly, \citet{Li2024} identified variable QPOs at 0.14 Hz, 0.18 Hz, 0.21 Hz, and 0.3 Hz using \insighthxmt{} data, which also appeared in the supercritical regime. Additionally, they observed excess power in the power spectra, which was modelled by adding another Lorentzian component with centroid frequencies around 0.37--0.48 Hz. 

The source was also followed up using \astrosat{} during different phases of the outburst. The broadband coverage of \astrosat{} allowed us to explore the source properties in the wide energy range by combining data from SXT, LAXPC, and CZTI. In this paper, we utilized \astrosat{} observations to explore the evolution of several spectral and timing properties of \src. The structure of the paper is as follows: Section \ref{obs} describes the data reduction and analysis techniques. The timing (pulse and quasi-periodic) and spectral analysis results are summarized in Section \ref{res}. The results are discussed and concluded in Section \ref{dis}. % Section \ref{con} summarizes the conclusions of the study.

\section{Observation and data analysis}
\label{obs}

\begin{table*}
    \centering
    \caption{The log of \astrosat\ observations of \src\ analyzed in this  work. The spin period and luminosity value quoted are from sec \ref{timing} and \ref{spectral}.} 
    \resizebox{1.4\columnwidth}{!}{
    \begin{tabular}{ccccccc}
    \hline
     Obs & Obs-ID    & \multicolumn{2}{c}{Date}          & Exposure  &  Luminosity  & Spin Period \\
      No.   &       & (yy-mm-dd) & MJD   & (ks)      &      (\lum)         & (s)    \\
    \hline  
     1 &  T05\_074T01\_9000005472 & 2023-01-11 & 59955 & 22 & $0.33 \times 10^{37}$&   207.789 (1) \\
     2 &  T05\_086T01\_9000005496  & 2023-02-05 & 59980 & 46 & $3.3 \times 10^{37}$&  206.5118 (2)  \\
     3 &  T05\_086T01\_9000005518  & 2023-02-25 & 60000 & 53 & $2.6 \times 10^{37}$&   205.3518 (1)  \\
     4 &  T05\_086T01\_9000005534 & 2023-03-07 & 60010 & 51 & $1.5 \times 10^{37}$&   204.9997 (1) \\
    \hline     
    \end{tabular}}    
    \label{tab:obs}
\end{table*}

\begin{figure}
\centering
 \includegraphics[width=0.9\linewidth]{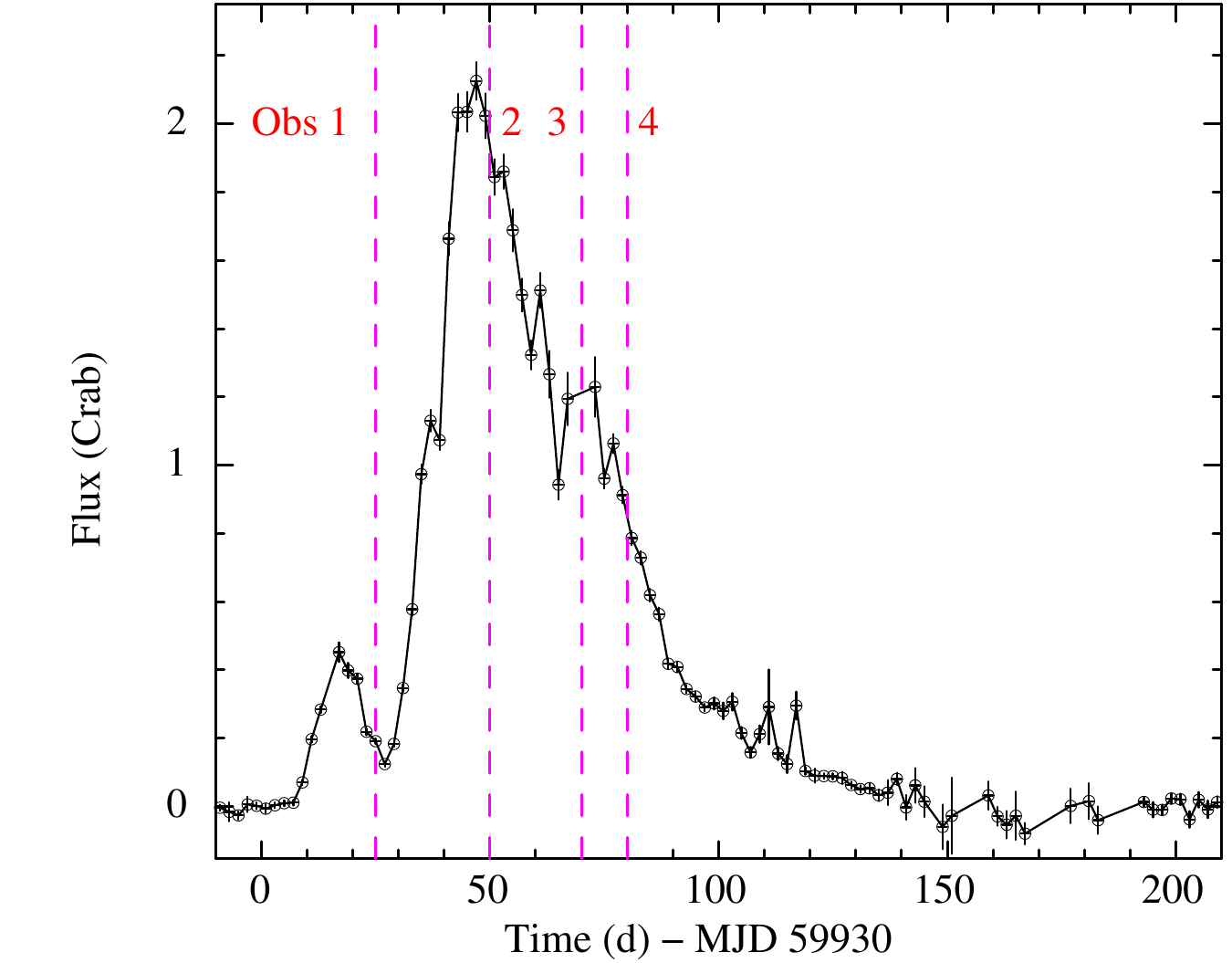}
 \caption{The \swiftbat{} lightcurve of \src\ during its 2022--2023 outburst binned at 2 days. The vertical magenta lines represent the epoch of \astrosat\ observations.}
\label{fig:outburst-lc}
\end{figure}

\astrosat\ is India's first dedicated multi-wavelength astronomy satellite \citep{Agrawal2006, Singh2014}, launched in 2015. It has five principal payloads on board: (i) the Soft X-ray Telescope (SXT), (ii) the Large Area X-ray Proportional Counters (LAXPCs), (iii) the Cadmium-Zinc-Telluride Imager (CZTI), (iv) the Ultra-Violet Imaging Telescope (UVIT), and (v) the Scanning Sky Monitor (SSM). Table~\ref{tab:obs} gives the log of observations used in this work. We analyzed data from SXT, LAXPC, and CZTI instruments.

\subsection{LAXPC}

LAXPC is one of the primary instruments aboard \astrosat. It consists of three co-aligned identical proportional counters (LAXPC10, LAXPC20, and LAXPC30) that work in the energy range of 3--80 keV. Each LAXPC detector independently records the arrival time of each photon with a time resolution of $10 ~\mu$s and has five layers \citep[for details see][]{Yadav2016, Antia2017}.
As LAXPC10 was operating at low gain and detector LAXPC30\footnote{LAXPC30 has been switched off since 8 March 2018 due to abnormal gain changes; see, \url{http://astrosat-ssc.iucaa.in/}} was switched off during the observation, we used only the LAXPC20 detector for our analysis. We used the data collected in the Event Analysis (EA) mode and processed using the \textsc{LaxpcSoft}\footnote{\url{http://www.tifr.res.in/~astrosat\_laxpc/LaxpcSoft.html}} version 3.4.4 software package to extract light curves and spectra.  LAXPC detectors have a dead-time of $42~\mu$s, and the extracted products are dead-time corrected.
The background in LAXPC is estimated from the blank sky observations \citep[see][for details]{Antia2017}. We have used corresponding response files to obtain channel-to-energy conversion information while performing energy-resolved analysis.

We corrected the LAXPC photon arrival times to the Solar system barycentre by using the  \textsc{as1bary}\footnote{\url{http://astrosat-ssc.iucaa.in/?q=data\_and\_analysis}} tool. We used the best available position of the source, R.A. (J2000)$=04^h 40^m 59.33^s$ and Dec. (J2000) $=44^{\circ} 31' 49.258''$ \citep{Gaia}.

\subsection{SXT}

The Soft X-ray Telescope (SXT) is a focusing X-ray telescope with CCD in the focal plane that can perform X-ray imaging and spectroscopy in the 0.3--7 keV energy range \citep{Singh2016, Singh2017}. 
\src\ was observed in the Photon Counting (PC) mode with SXT (Table~\ref{tab:obs}). Level 1 data were processed with \texttt{AS1SXTLevel2-1.4b} pipeline software to produce level 2 clean event files. The level 2 cleaned files from individual orbits were merged using the SXT event merger tool (Julia Code). The merged event file was then used to extract images, light curves, and spectra using the \textsc{xselect} task, provided as part of \textsc{heasoft} version 6.31.1. A circular region with a radius of 16 arcmin centred on the source was used.
For observations 1, 3, and 4, the source count rate was below the threshold value of pileup ($<40$ counts \psec) in the PC mode\footnote{\url{https://www.tifr.res.in/~astrosat_sxt/instrument.html}}. For observation 2, the source count rate was significantly above the threshold of pileup ($>40$ counts \psec). To mitigate the effect of the pileup, the source counts were excluded from the central region of the point-spread function (PSF) until the source count rate became below the threshold rate for the pileup. An annulus region with an inner radius of $2.5$ arcmin and an outer radius of 16 arcmin was chosen. We also checked for pile-up effects in these observations using the method described by \citet{Sridhar2019, Chakraborty2020}.
For spectral analysis, we have used the blank sky SXT spectrum as background (SkyBkg\_sxt\_LE0p35\_R16p0\_v05\_Gd0to12.pha) and spectral redistribution matrix file (sxt\_pc\_mat\_g0to12.rmf) provided by the SXT team\footnote{\url{http://www.tifr.res.in/~astrosat\_sxt/dataanalysis.html}}. We generated the correct off-axis auxiliary response files (ARF) using the sxtARFModule tool from the on-axis ARF (sxt\_pc\_excl00\_v04\_20190608.arf) provided by the SXT instrument team.

\subsection{CZTI}

The Cadmium Zinc Telluride Imager (CZTI) is a hard X-ray instrument operating in the 20--200 keV energy range with a coded mask aperture to subtract the background \citep{Bha17}. It has four independent quadrants, each with 16 pixelated detector modules. CZTI provides an angular resolution of 8 arcmin with a field of view of 4.6$^{\circ}$ $\times$ 4.6$^{\circ}$. The CZTI data were reduced using the CZTI data analysis pipeline version 3.0. We used the \texttt{cztpipeline} tool to generate level-2 cleaned event files from the level-1 data. The spectrum and light curves were extracted using \texttt{cztbindata}. The spectrum from all four quadrants of CZTI was added using \texttt{cztaddspec}. The light curves for different energy ranges were binned at 1 s time bins and barycenter corrected using \textsc{as1bary} tool. The spectra are used in the energy range of 20--150 keV for Obs. 2, 3, and 4. For Obs. 1, we have used the CZTI lightcurves and spectra for the energy range of 20--50 keV. The final data products (light curves and spectra) are created for the mentioned energy ranges, beyond which the source has poor count statistics.

%%%%%%%%%%%%%%%%%%%%%%%%%%%%%%%%%%%%%%%%%%%%%%%%%%%%%%%%%%%%%%%%%%%%%%%%%%%%%%%%%%%%%%%%%%%%%%%%
%%%%%%%%%%%%%%%%%%%%%%%%%%%%%%%%%%%%%%%%%%%%%%%%%%%%%%%%%%%%%%%%%%%%%%%%%%%%%%%%%%%%%%%%%%%%%%%%
%%%%%%%%%%%%%%%%%%%%%%%%%%%%%%%%%%%%%%%%%%%%%%%%%%%%%%%%%%%%%%%%%%%%%%%%%%%%%%%%%%%%%%%%%%%%%%%%

\section{Results}
\label{res}

\begin{figure}
\centering
 \includegraphics[width=0.9\linewidth]{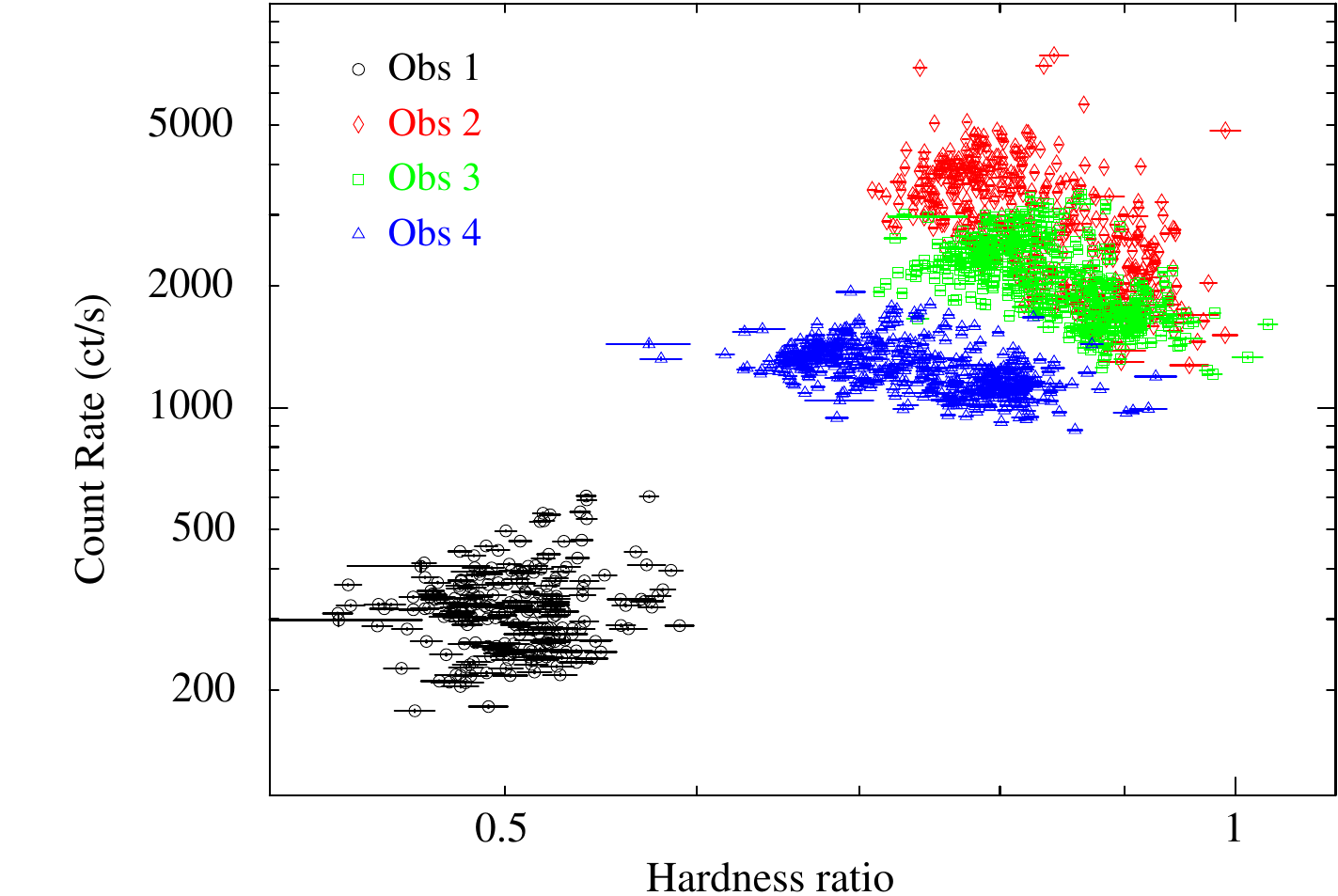}
 \caption{The hardness-intensity diagram of \src\ using LAXPC. The count rate is given in the energy range of 3--30 keV, and the hardness ratio is the ratio of count rate in the energy range of 10--30 and 3--10 keV.}
\label{fig:hid}
\end{figure}

Figure \ref{fig:outburst-lc} presents the 15--50 keV \swift/BAT light curve \citep{Krimm2013} of the 2022--2023 giant outburst of \src, where the peak flux reached a record-high value of $\sim$2.3 Crab near MJD 59976. \astrosat\ observed the source at four epochs at different phases of the outburst and luminosity, marked with vertical dashed lines in Fig.  \ref{fig:outburst-lc}.  

The evolution of the hardness ratio of \src{} with intensity during these \astrosat{} observations is shown in Fig. \ref{fig:hid}. The hardness ratio is estimated from the ratio of the count rate in the energy band 10--30 keV and 3--10 keV of LAXPC. A significant variation in the hardness ratio with the source intensity is observed, indicating different phases of outburst. A detailed study using \nicer{} observations traced the source state transition of the \src\ during the giant outburst in 2022, indicating a state transition from subcritical to supercritical regime occurred at MJD 59969–59970 and opposite transition on MJD 59995 \citep{Mandal23, Sa23}. 

In our study, the \astrosat{} observation close to the peak belongs to the supercritical state of the source, while other observations were performed in the subcritical state of the source \citep{Mandal23, Sa23}. A significant evolution in temporal and spectral properties is expected between these observations. The \astrosat{} observations allowed us to probe the broadband view of the source at different accretion regimes.

%%%%%%%%%%%%%%%%%%%%%%%%%%%%%%%%%%%%%%%%%%%%%%%%%%%%%%%%%%%%%%%%%%%%%%%%%%%%%%%%%
%%%%%%%%%%%%%%%%%%%%%%%%%%%%%%%%%%%%%%%%%%%%%%%%%%%%%%%%%%%%%%%%%%%%%%%%%%%%%%%%%
\begin{figure*}
\centering
 \includegraphics[width=0.4\linewidth]{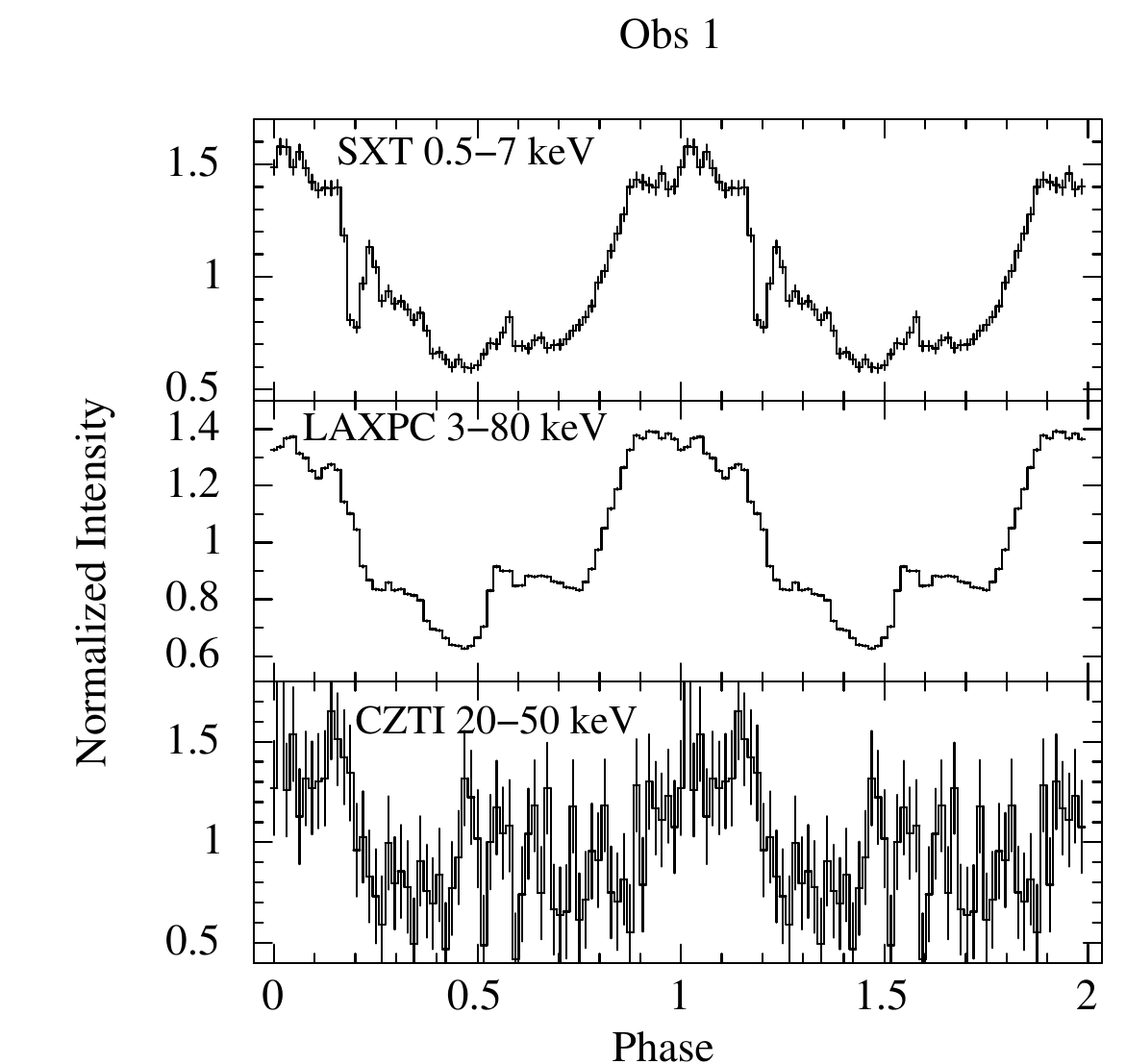}\vspace{0.5cm}
  \includegraphics[width=0.4\linewidth]{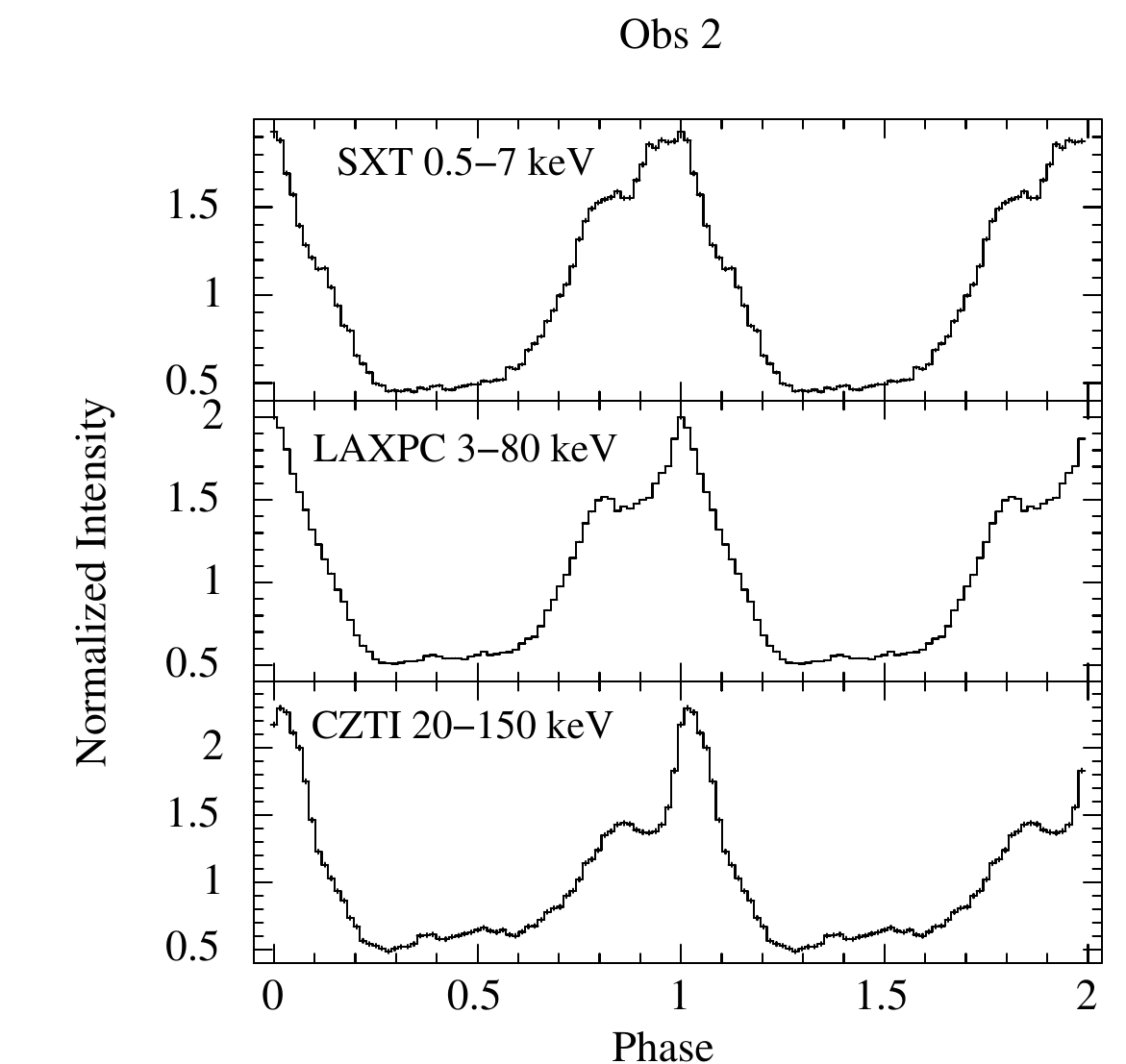}
   \includegraphics[width=0.4\linewidth]{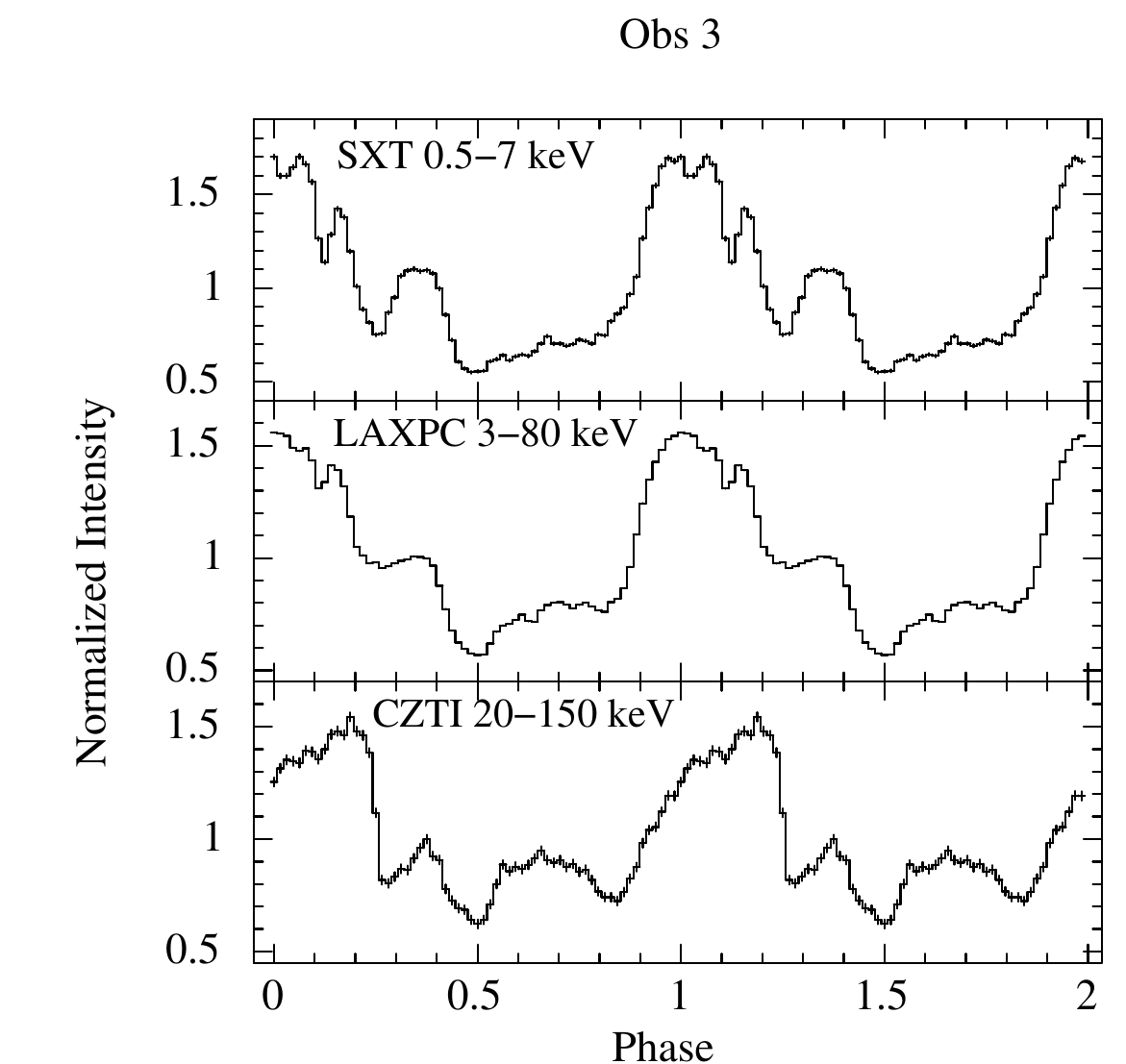}
    \includegraphics[width=0.4\linewidth]{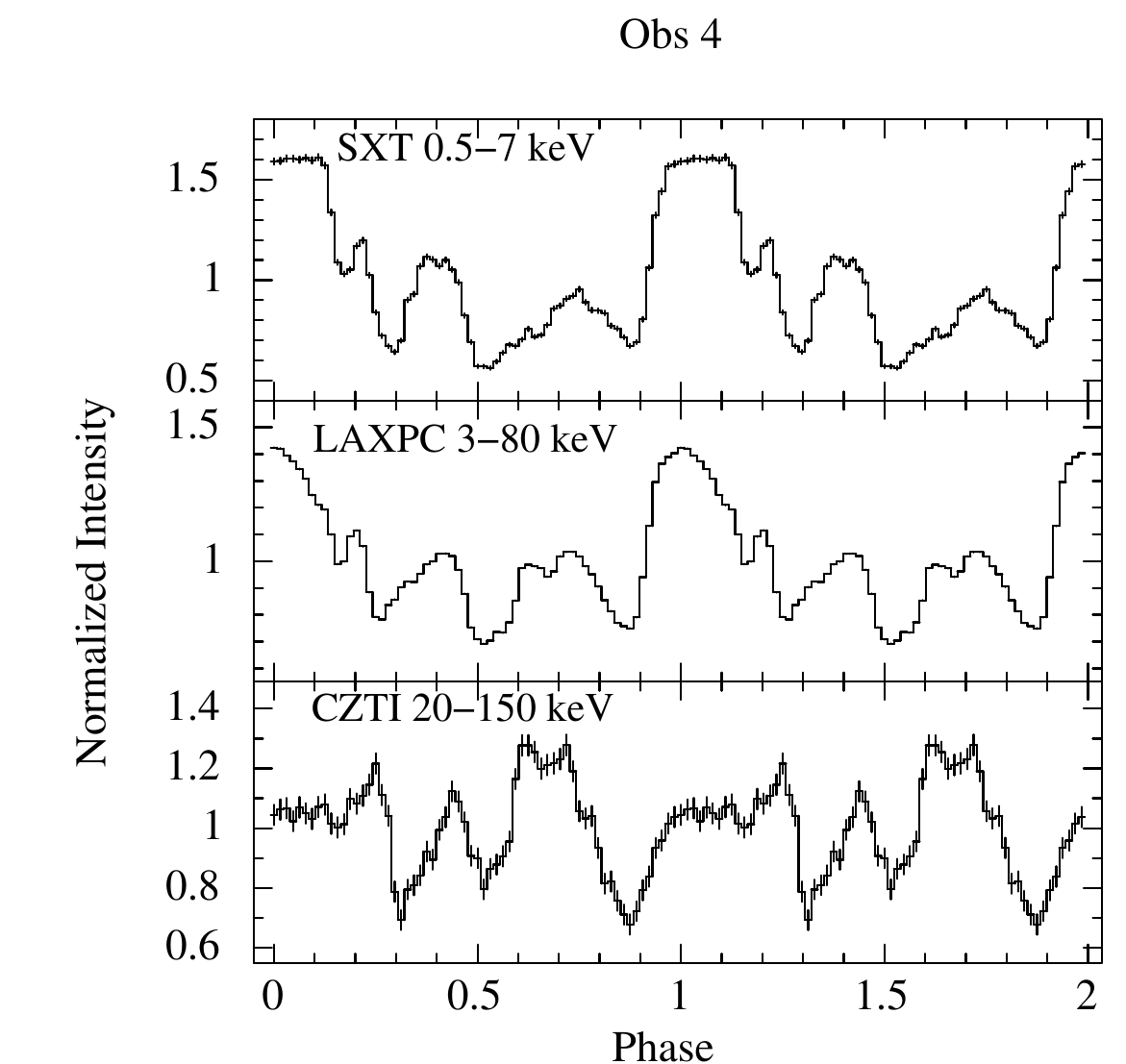}
\caption{The pulse profiles of \src{} from \astrosat--SXT, LAXPC, and CZTI observations during different phases of the outburst. Pulsations in the observation were detected up to 50 keV, while for observations 2, 3, and 4, pulsations were detected up to 150 keV.}
    \label{fig:pp-all}
\end{figure*}

\begin{figure*}
\centering
 \includegraphics[width=0.24\linewidth]{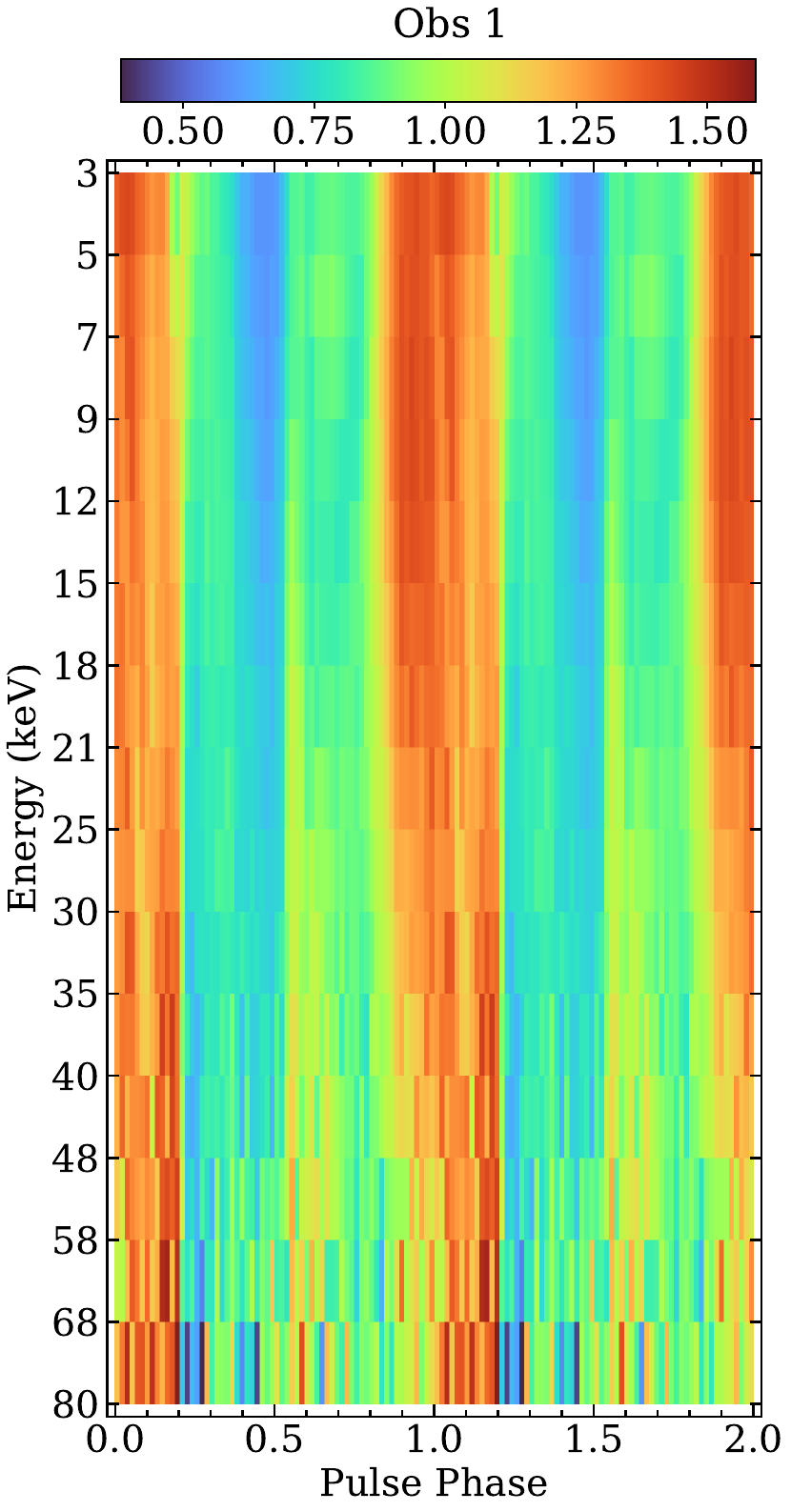}
  \includegraphics[width=0.24\linewidth]{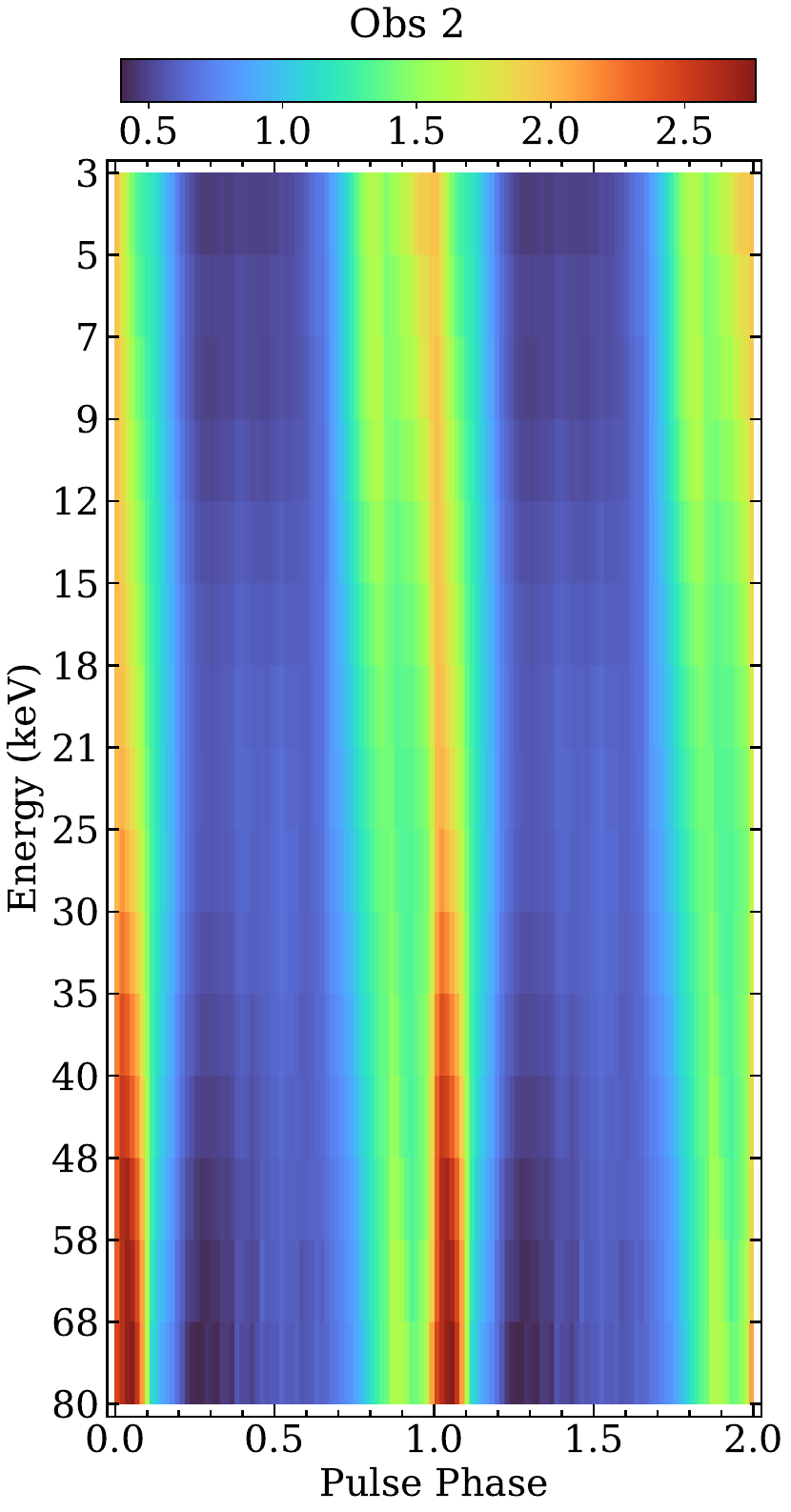}
   \includegraphics[width=0.24\linewidth]{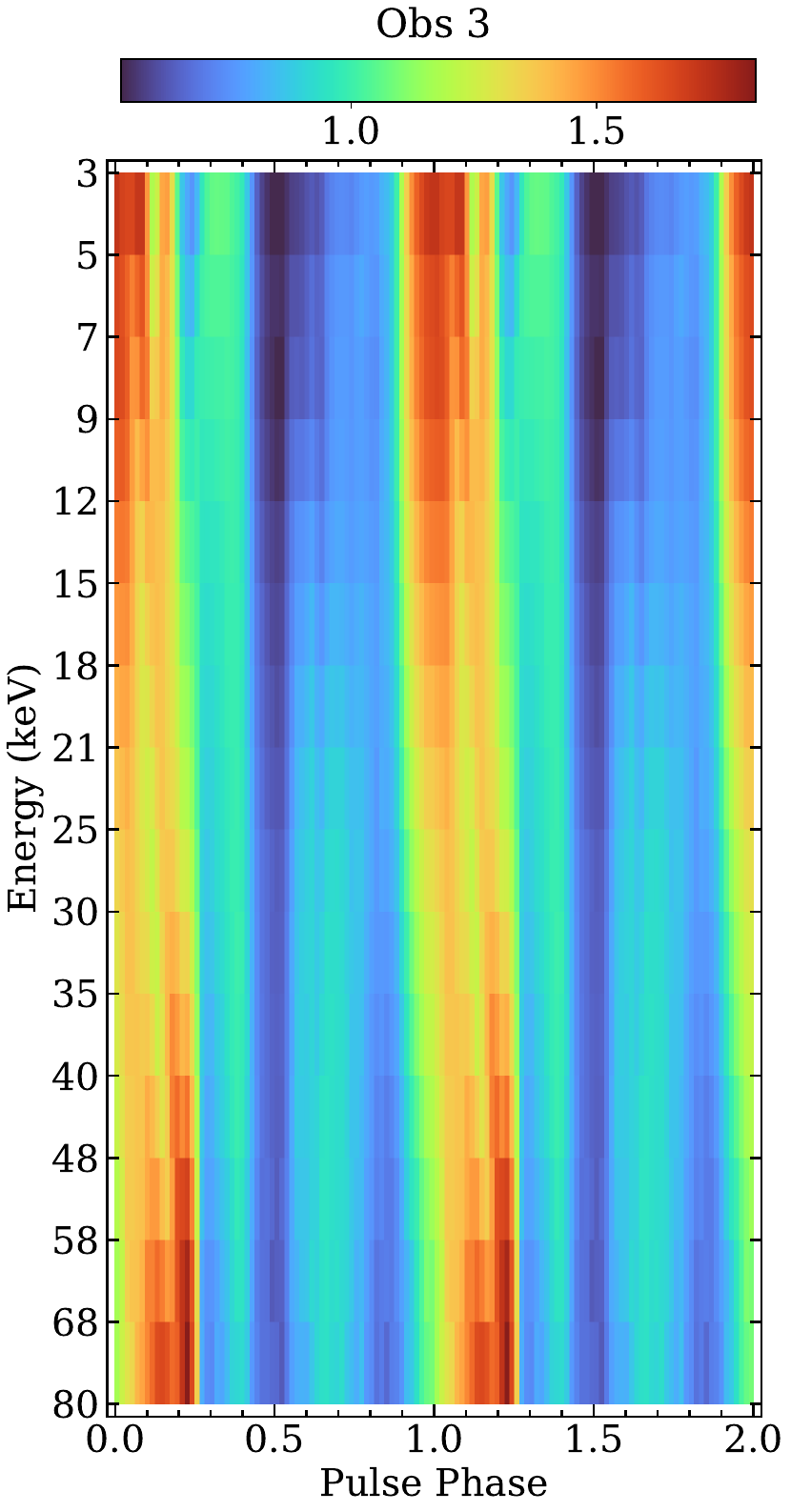}
    \includegraphics[width=0.24\linewidth]{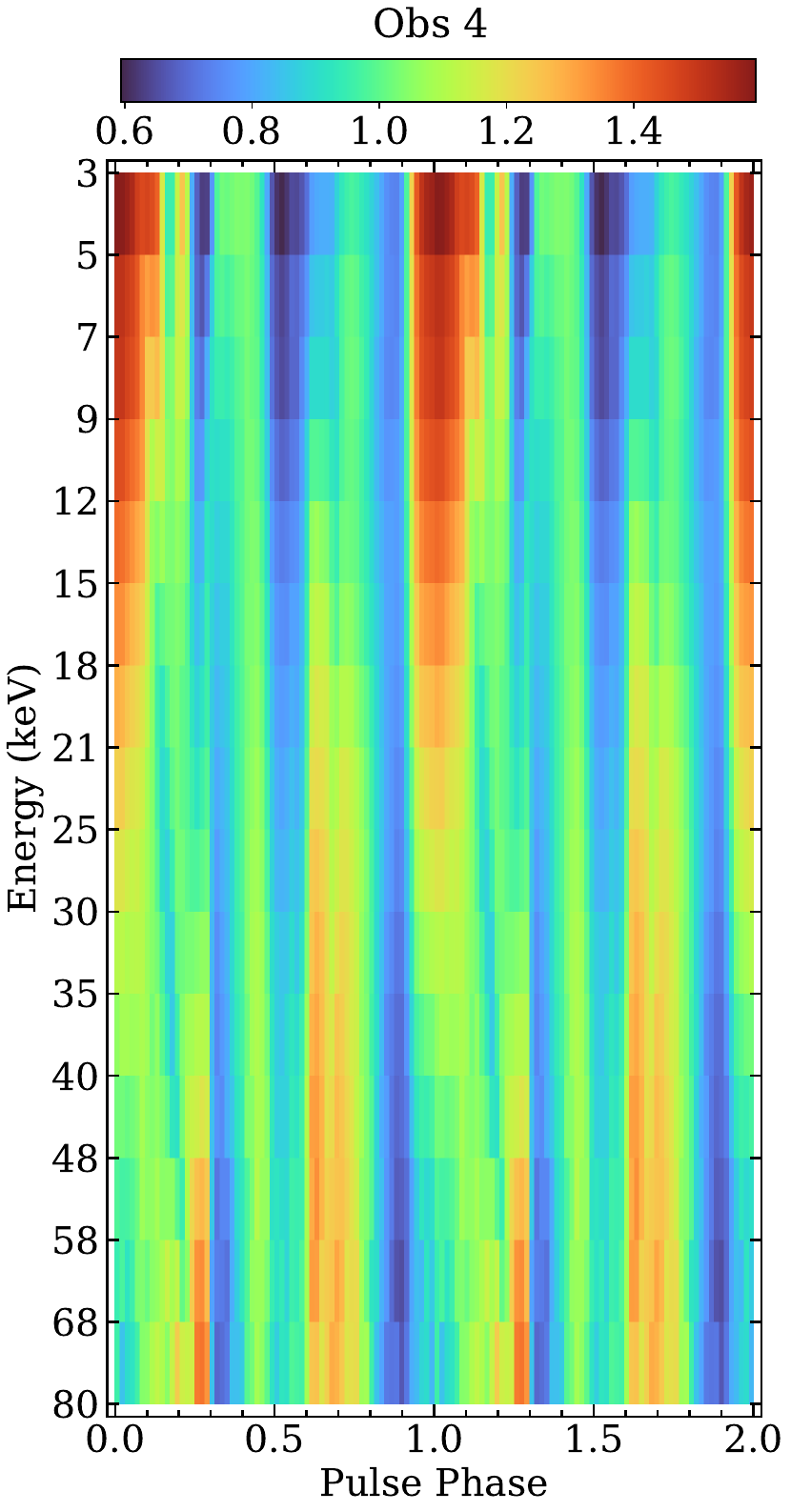}
 \caption{Colormap of the Energy dependence of the pulse profiles from four \astrosat{}/LAXPC observations at different luminosity levels of the outburst. The color bar on top represents the normalized intensity of pulse profiles.}
\label{fig:pp2}
\end{figure*}

\begin{figure}
\centering
  \includegraphics[width=0.9\linewidth]{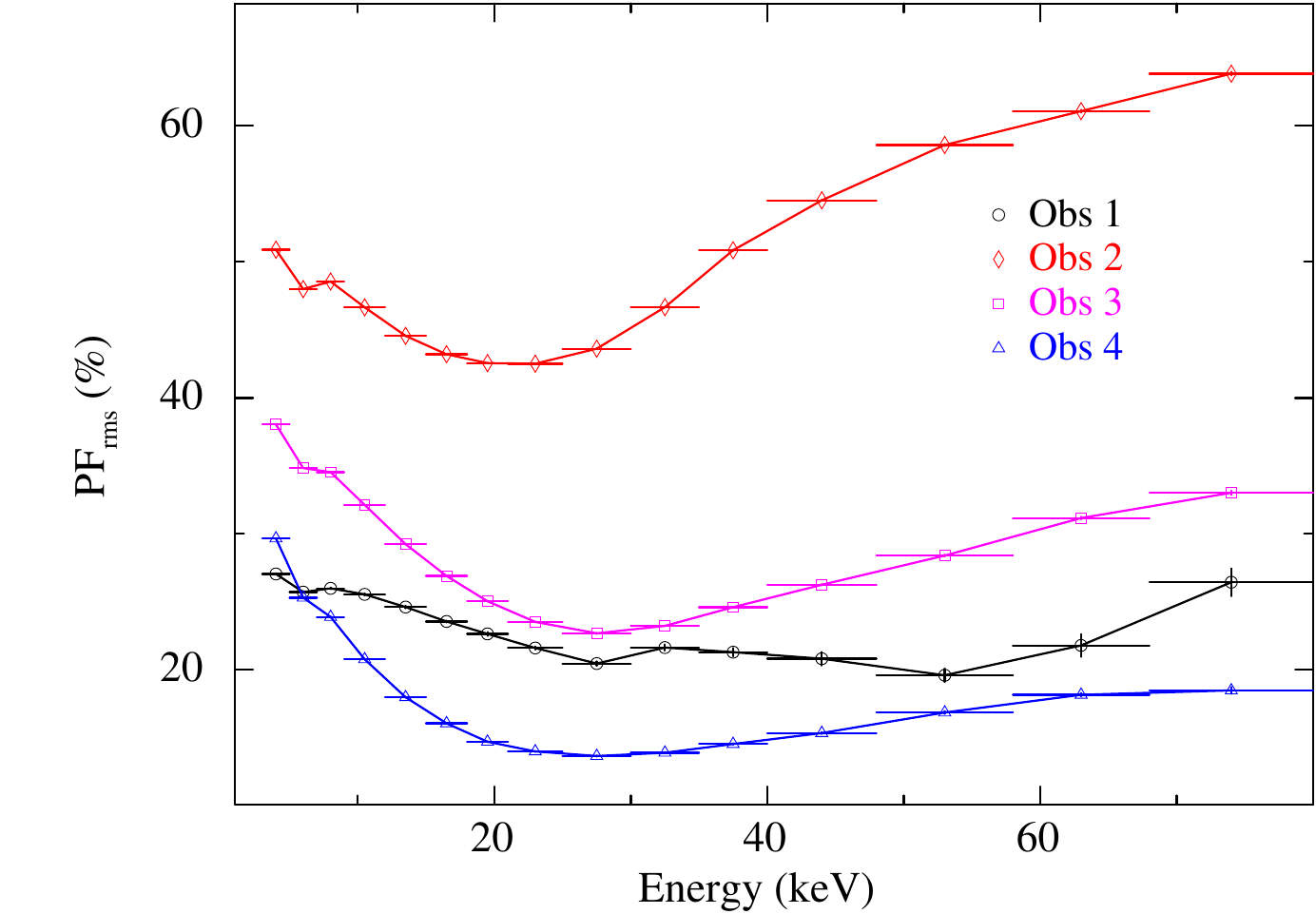}
 \caption{Variation of rms pulsed fraction with energy from all four \astrosat{} LAXPC observations. }
\label{fig:pf}
\end{figure}

\subsection{Pulse timing analysis}
\label{timing}

The 1-s binned background-corrected light curves from LAXPC in the 3--80 keV energy range were used to search for periodicity from the \astrosat{} observations during the outburst. The epoch-folding technique was employed by using the {\tt efsearch} task to find the optimal period by folding the light curve over a trial period range and maximizing $\chi{^2}$ \citep{Leahy87} as a function of the period across 32 phase bins. The bootstrap method \citep{Lutovinov12} was used to determine the uncertainty in the estimated spin periods by simulating 1000 light curves, as explained in \citet{Sharma23}. During the \astrosat{} observations, the spin period of \src{} varied between 207.789 s to 204.9997 s; details are summarised in Table \ref{tab:obs}. The spin-period is consistent with the \fermigbm{} \citep{Malacaria2020} and \nicer{} measurements \citep{Mandal23}, and the spin-up of \src{} is evident during the outburst.

The determining optimal spin periods were used to produce pulse profiles by folding light curves from SXT, LAXPC, and CZTI instruments using the {\tt efold} task. During the \astrosat{} observations, the source also exhibited significant changes in pulse profile shape. The pulse profile in 0.5--150 keV is shown in Fig. \ref{fig:pp-all} for different \astrosat{} observations during the outburst. The SXT (0.5--7 keV), LAXPC (3--80 keV), and CZTI (20--150 keV) pulse profiles are shown in the top, middle, and bottom panels of Fig. \ref{fig:pp-all}. For observation 1, the CZTI pulse profile was detected up to 50 keV due to poor source count at high energy. The multi-peak feature is observed at lower luminosity levels with an additional dip-like feature at low energies \citep{Pa12, Ts12, Usui2012, Mandal23}, whereas a single asymmetric broad peak feature with a minor peak \citep[or two wings with phase separation of $\sim$0.2;][]{Li23, Sa23} is visible in the pulse profiles (during Obs. 2) at the peak of the outburst. During the decay phase of the outburst (Obs 3 and 4), the pulse profile evolves back to a multi-peaked feature with more complexity. A similar behaviour of pulse profiles at different luminosities was reported previously \citep{Li23, Mandal23, Sa23}.
  
Fig. \ref{fig:pp2} shows the energy-dependence of pulse profiles from four different \astrosat{} observations from LAXPC in the 3--80 keV range at different luminosities of the outburst (Fig. \ref{apx:pp} present the individual energy-resolved pulse profiles). Strong energy dependence was evident in the pulse profiles. In Obs 1, the pulse profile shows a highly energy-dependent complex dual-peak feature. The main peak ($\sim$1 phase) significantly evolved with energy. The right wing of the main peak grows sharply with an increase in energy and becomes a more prominent feature above 40 keV, a narrower peak at the $\sim1.2$ phase. 
At the peak of the outburst (during Obs 2), the pulse profile evolved into a single broad peak with a minor peak feature before the main peak, which was strongly energy-dependent. The main peak showed a sharp increase in intensity with increasing energy. The main and minor peaks shifted to the right, and their separation decreased with increasing energy. 

With the decay of the outburst (Obs 3 and Obs 4), the pulse profile evolved back to a multi-peak nature with a strong energy dependency and dip-like feature at low energies. During Obs 3, the pulse peak shifted to the right, with minima remaining at phase $\sim$0.5. A minor peak around 0.6--0.7 phase becomes apparent at higher energies. In middle energies (15--48 keV), a dip-like feature also appears on the main peak and separates it into wings. This dip shifts from phase 0.1 to 0.2 with the pulse peak. Similar to Obs 1, the right wing of the peak evolved into a sharp peak at phase $\sim1.2$ at higher energies.

The pulse profile was more complex during Obs 4, with three peaks and a dip-like feature in the right wing of the main peak. The contribution significantly changed for these peaks with energies. The dominance of the peak from main (at phase$\sim$1) to secondary (at phase$\sim$0.7) changed above $\sim$25 keV. In this case, the right wing of the main peak (at phase $\sim$1)  diminished with an increase of energy and started to evolve into a sharp peak at phase $\sim$0.35 above 48 keV, similar to Obs 1 and 3.

We also studied the energy-resolved pulse profiles from CZTI in the energies up to 150 keV, shown in Fig. \ref{apx:pp_CZTI}. The pulse profiles are shown for the energy ranges 20--50 keV, 50--100 keV, and 100--150 keV. At higher luminosity (Obs 2), the pulse profile showed a broad single peak feature with a minor peak in 20--150 keV, similar to LAXPC. In the decay phase of the outburst (Obs 3 and Obs 4), the pulse profile exhibited a multi-peak feature which evolved significantly with energy and luminosity. Above 100 keV, only the primary peak stands out, while minor peaks are not clear. 

\subsubsection{Pulsed fraction}

The variation of the rms pulsed fraction (PF) with energy is also studied during different \astrosat{} observations. The rms PF is estimated using the following \citep{Wilson2018}, 
\begin{equation}
    {\mathrm{PF_{rms}}} = \frac{1}{\bar{p} \sqrt{N}} \left[\sum \limits_{i=1}^N (p_{i} -\bar{p})^2\right]^\frac{1}{2}
\end{equation}
where $N$ is the number of phase bins, $\bar{p}$ is the average count rate, and $p_{i}$ is the count rate in the {\it i}$^{\rm th}$ phase bin of the pulse profile.
The PFs changed significantly during the outburst and showed strong luminosity and energy dependence. The variation of the PF with energy during different \astrosat{} observations is shown in Fig. \ref{fig:pf}. The PFs are found to be higher close to the peak of the outburst and as the outburst evolved, the PF showed comparatively low value. 

The PF shows a negative correlation below the energy of 25 keV, and the correlation turns positive above 25 keV \citep{Li23, Sa23}. In Obs 1, the PF decreased from $\sim$27\% to 20\% below 30 keV, and above this energy range, the PF increased back to 26\% at 68--80 keV. Near the peak of the outburst (Obs 2), the PF decreased from $\sim$50\% to $\sim$42\% below 25 keV and showed a sharp increase up to $\sim$64\% above 25 keV. In Obs 3, the PF decreases from $\sim$38\% to 22\% in the 3--30 keV energy range, and afterward, the PF increases up to $\sim$33\% at 68--80 keV. In Obs 4, the PF decreased from $\sim$30\% to 13\% below 30 keV and increased to $\sim$18\% afterward. During all \astrosat{} observations, the PF showed a minor dip close to the energy 6--7 keV, similar to \nustar{} observations \citep{Sa23}. This local minimum feature in PF is evident around the fluorescence iron line. A concave-like feature is evident in the PF plots around 20--30 keV, probably due to the scattering of high energy photons in the neutron star atmosphere, which reduces the amplitude of the pulsed emission \citep{Li23}.

%%%%%%%%%%%%%%%%%%%%%%%%%%%%%%%%%%%%%%%%%%%%%%%%%%%%%%%%%%%%%%%%%%%%%%%%
%%%%%%%%%%%%%%%%%%%%%%%%%%%%%%%%%%%%%%%%%%%%%%%%%%%%%%%%%%%%%%%%%%%%%%%%
%%%%%%%%%%%%%%%%%%%%%%%%%%%%%%%%%%%%%%%%%%%%%%%%%%%%%%%%%%%%%%%%%%%%%%%%

\subsection{Quasi-periodic variability}
\label{qpo}

\begin{figure*}
\centering
 \includegraphics[width=0.49\linewidth]{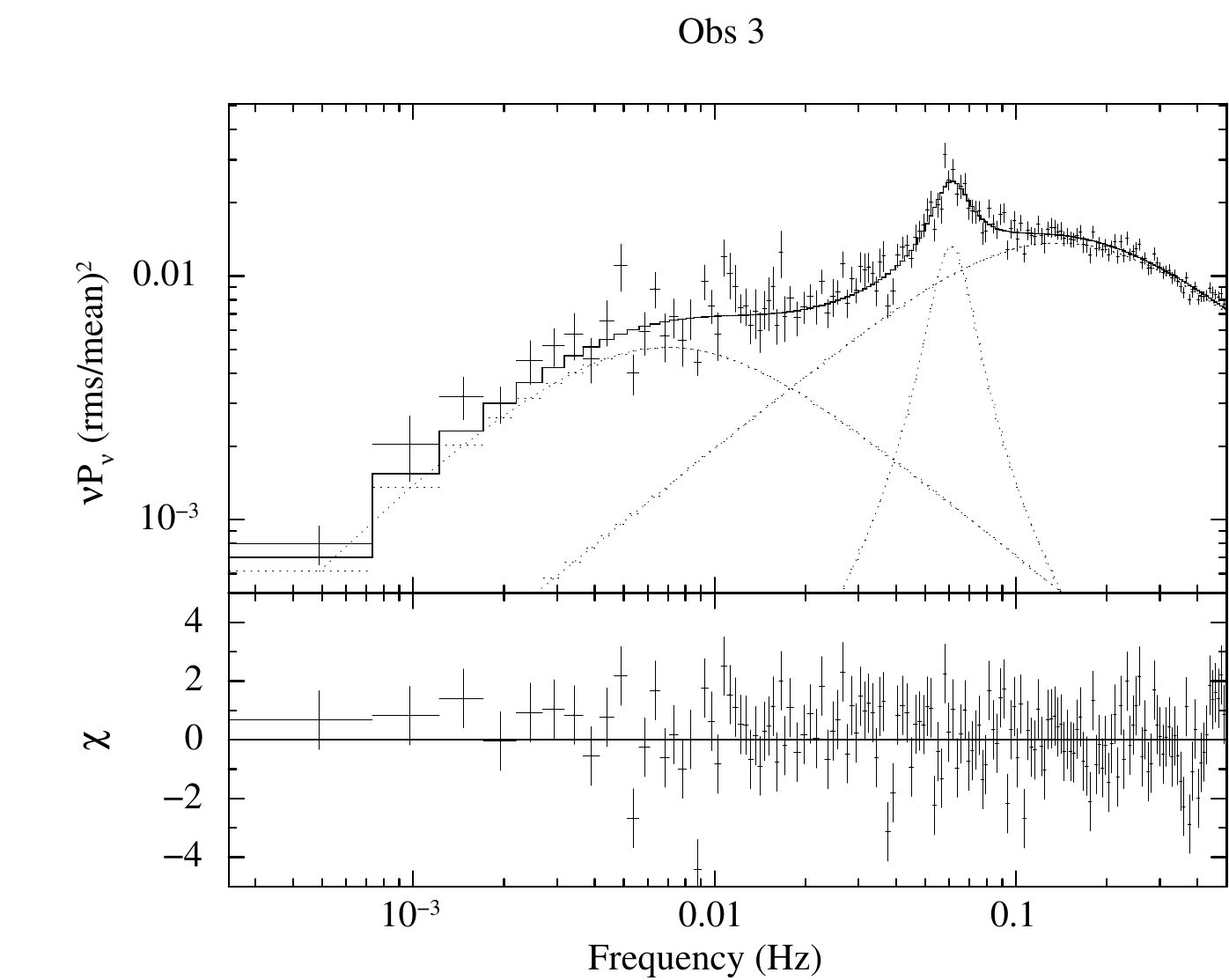}
  \includegraphics[width=0.49\linewidth]{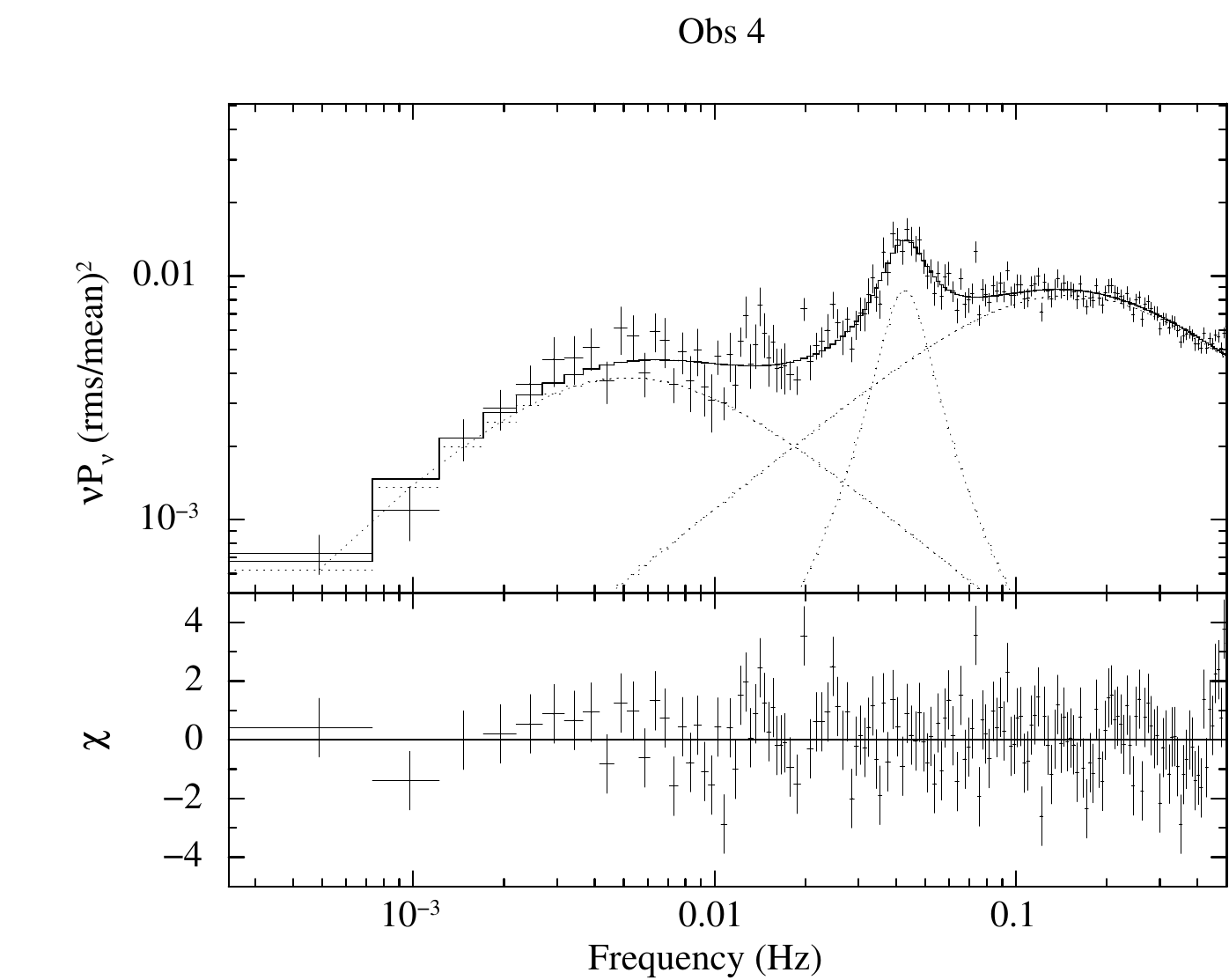}
 \caption{The 3--80 keV PDS of \src\ from LAXPC data of observation 3 (left) and 4 (right). PDS can be modelled with two Loretzian functions for the noise continuum and an additional Loretzian to model the QPO at 60 mHz and 42 mHz for observations 3 and 4, respectively. A thick solid line in the plot indicates the best-fit model, while the dotted lines represent the individual components of the model.}
\label{fig:qpo}
\end{figure*}

\begin{figure*}
\centering
 \includegraphics[width=0.459\linewidth]{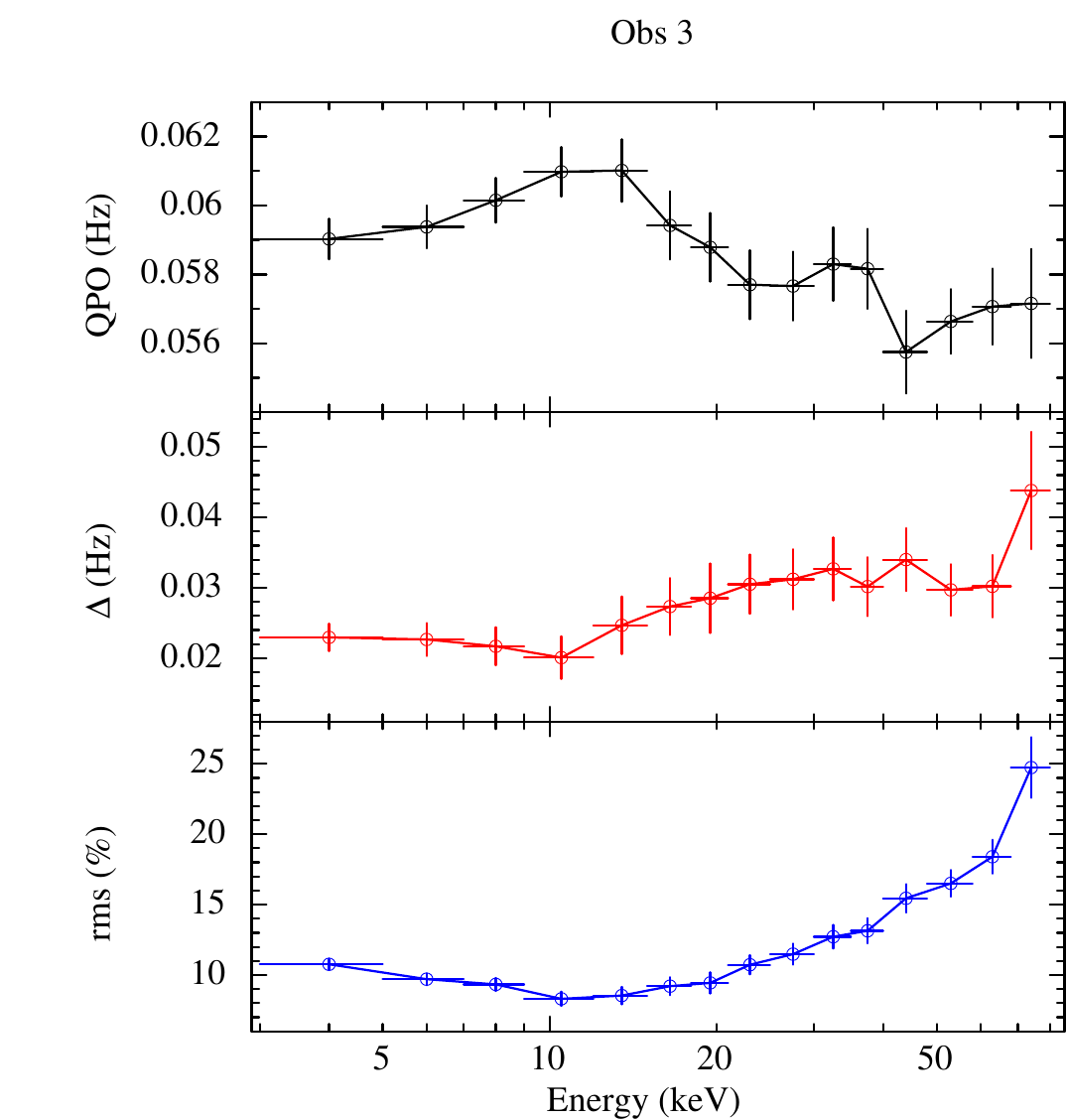}
  \includegraphics[width=0.459\linewidth]{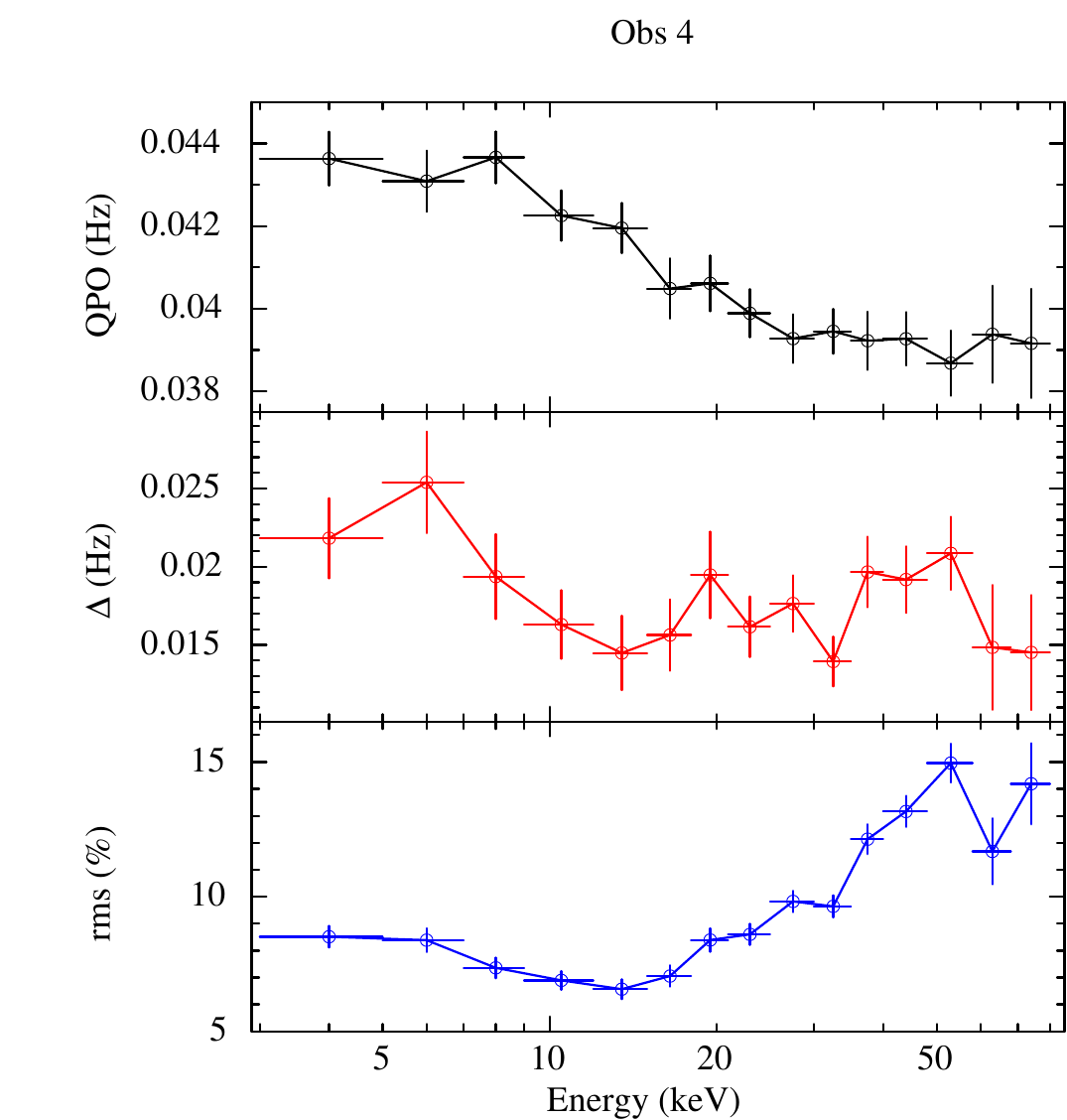}
 \caption{Variation of the QPO parameters with energy for the LAXPC observations 3 (left) and 4 (right).}
\label{fig:e-qpo}
\end{figure*}

\begin{table}
	\centering
	\caption{The best-fit parameters of the PDS for  the \astrosat{} observation 3 and 4. Reported errors are at a 90\% confidence level for a single parameter.}
	\resizebox{0.9\linewidth}{!}{
\begin{tabular}{c c c c}
\hline
Component & Parameter &  Obs 3 & Obs 4 \\
\hline
Lore 1 & Freq., $\nu_0$ (Hz) & $0^{\rm fixed}$ & $0^{\rm fixed}$ \\[1ex]
& FWHM, $\Delta$ (Hz) & $0.014 \pm 0.002$ &  $0.0103_{-0.0011}^{+0.0014}$ \\[1ex]
&rms (\%) & $17.9 \pm 0.7$ &  $15.5 \pm 0.6$ \\[1ex]

Lore 2 & Freq., $\nu_0$ (Hz) & $0^{\rm fixed}$& $0^{\rm fixed}$ \\[1ex]
&FWHM, $\Delta$ (Hz) & $0.278 \pm 0.015$ & $0.300 \pm 0.012$\\[1ex]
&rms (\%) & $29.3 \pm 0.6$ & $22.8 \pm 0.3$\\[1ex]

Lore 3 & Freq., $\nu_0$ (Hz) & $0.060 \pm 0.001$ & $0.042 \pm 0.001$  \\[1ex]
&FWHM, $\Delta$ (Hz) & $0.021 \pm 0.005$  & $0.017_{-0.003}^{+0.004}$ \\[1ex]
&rms (\%) & $8.5 \pm 0.8$ &  $7.5 \pm 0.5$ \\[1ex]
&Quality Factor, $Q$ & 2.8 & 2.4 \\   
& Significance ($\sigma$) & 17 & 25 \\
&$\chi^2/{\rm dof}$ &  234/151 & 233/151 \\

\hline
\end{tabular}}
\label{tab:pds}
\end{table}

To investigate variability, we constructed the power density spectrum (PDS) using LAXPC data in the 3--80 keV energy range across all four observations. The LAXPC light curves were binned at 0.1 sec and divided into $\sim$819.2 sec segments to calculate the Fourier transform. The resulting power spectra were averaged and rebinned geometrically by a factor of 1.02. The power spectra were calculated using rms normalization, with Poisson noise subtracted using \textsc{ftool} \texttt{powspec norm = --2}. Each PDS exhibited narrow peaks at the spin frequency of the pulsar ($\sim$5 mHz) with their multiple harmonics and red noise. Additionally, the PDS from observations 3 and 4 showed a narrow excess in the power around 60 mHz and 42 mHz, respectively, indicating a quasi-periodic oscillation (QPO). 

The aperiodic noise in accreting X-ray pulsar can be explained by the propagating fluctuations in mass accretion rate occurring in the accretion disc \citep[e.g.,][]{Lyubarskii1997, Churazov2001, Revnivtsev2009}, and PDS can be described using a characteristic power-law shape with a break. The break frequency is associated with the change of the disc-like accretion flow to the magnetospheric flow influenced by the interaction between the accretion disc and the compact object, the role of magnetic fields and mass accretion rate \citep{Revnivtsev2009, Mushtukov2019}. The PDS can also be described by a sum of Lorentzian functions \citep[e.g.,][]{Belloni2002, Reig2008}. The Lorentzian profile is a function of frequency and is defined as
   \begin{eqnarray}
   \label{eqn:lore}
   P(\nu) = \frac{r^2 \Delta}{2 \pi} \frac{1}{(\nu - \nu_0)^2 + (\Delta/2)^2},
   \end{eqnarray}
where $\nu_0$ is the centroid frequency, $\Delta$ is the full-width at half-maximum (FWHM), and $r$ is the integrated fractional rms. The quality factor of Lorentzian $Q=\nu_0/\Delta$ is used to differentiate if a feature is a QPO or noise. The components with $Q>2$ are generally considered as QPOs, otherwise band-limited noise \citep[e.g.,][]{Belloni2002}. 

To accurately measure the fractional variability of the QPO, we corrected the light curve of observations 3 and 4 for spin modulations, as higher-order harmonic can interfere with QPO modelling. We modelled the mean pulse profile with a high-order harmonic Fourier function to remove the spin modulations \citep[see,][]{Finger1996}. 
Figure \ref{fig:qpo} shows the PDS of  \src\ from LAXPC20 in the 0.0005--0.5 Hz range, derived from the spin-modulation-corrected light curves of observations 3 (left) and 4 (right), respectively. 
The PDS showed a broad noise with excess power due to a QPO. The PDS continuum can be well-modelled with two broad Lorentzians centred at zero frequency. For broad noise components, the centroid frequency $\nu_0$ was close to zero, and $\Delta$ was adjusted to fit the characteristic frequency\footnote{In the $\nu \times P_\nu$ representation, the frequency at which maximum power is attained is the characteristic frequency, $\nu_{\rm max} = (\nu^2_0 + (\Delta/2)^2)^{1/2}$, rather than the centroid frequency of the Lorentzian, $\nu_0$.} of the feature \citep{Belloni2002}.
A third Lorentzian component was used to model the QPO feature. The QPOs exhibited a quality factor of $>$2 for both observations and rms of 8.5\% and 7.5\% for observations 3 and 4, respectively. Table \ref{tab:pds} summarizes the best-fit model parameters for QPOs. The significance of the QPOs was measured by taking the ratio between the normalization of the Lorentanzian function and its 1$\sigma$ error.

We also investigated the energy dependence of the QPOs. The PDS was extracted in narrower energy bands and modelled with a similar approach as the energy-average PDS. The variation of the QPO characteristics (centre frequency, FWHM, and rms amplitude) as a function of energy is shown in Figure \ref{fig:e-qpo}. The QPOs exhibited a clear energy dependence. For observation 3, the central frequency varied between $55.7 \pm 1.2$ mHz and $61.0 \pm 0.9$ mHz, and the FWHM increased with energy. Additionally, the QPO rms decreased up to $\sim$10 keV and then increased above this energy. In observation 4, QPO central frequency decreased with energy from $43.6 \pm 0.6$ mHz to $39.2 \pm 0.7$ mHz. The QPO rms showed a similar variation as of observation 3.

QPOs were also detected in the CZTI light curves. For Obs 3, a QPO was observed at $55.5 \pm 1.4$ mHz with an rms of 12.1$\pm$1.5\%, while for Obs 4, a QPO was detected at $40.5 \pm 1.8$ mHz with an rms of 15.5$\pm$2.9\% in the 20--150 keV energy band (Fig. \ref{fig:qpo-cs}). We also examined the energy-resolved light curve of CZTI to look for the evolution of QPO parameters. For Obs 3, the QPO was detected in the energy range of 20--50 and 50--100 keV, with rms increasing from 9.5$\pm$1.3\% to 23$\pm$2\% with increasing energy, consistent with LAXPC results. No QPO feature was detected above the 100 keV range. For obs 4, no QPO was significantly detected in the energy-resolved light curves.

Similarly, QPOs were detected in the SXT light curves (binned at the SXT readout time of 2.3775 s). In the 0.5--7 keV energy range, a QPO was observed at 58.8$ \pm $1.1 mHz with an rms of 8.6$ \pm $0.9\% for Obs 3, and 41.5$ \pm $1.5 mHz with an rms of 12.8$ \pm $2.5\% for Obs 4 (Fig. \ref{fig:qpo-cs}). For Obs 3, the QPO from the energy-resolved SXT light curves also showed energy dependence, consistent with LAXPC results. The QPO frequencies were 39.8$ \pm $2.2, 43.9$ \pm $1.8 and 46.2$ \pm $2.3 mHz with rms of 17.4$ \pm $3.6\%, 11.3$ \pm $1.8\% and 12.1$ \pm $2.4\% for the energy range of 0.5--2, 2--4 and 4--7 keV, respectively. For obs 4, no significant QPO was detected in the energy-resolved SXT light curves.

%%%%%%%%%%%%%%%%%%%%%%%%%%%%%%%%%%%%%%%%%%%%%%%%%%%%%%%%%%%%%%%%%%%%%%%%%%%%%%%%%%

\renewcommand{\arraystretch}{1.2}
\begin{table*}
	\centering
	\caption{Best-fit spectral parameters of \src{} with spectral models M1 (\texttt{tbabs$\times$(bbodyrad+bbodyrad+compTT+Gaussian)}) and M2 (\texttt{tbabs$\times$(bbodyrad+compTT+compTT+Gaussian)$\times$edge}). All errors and upper limits reported in this table are at a 90\% confidence level ($\Delta \chi^2=2.7$).}
	\label{tab:fitstat2}
	\resizebox{\linewidth}{!}{
	\begin{tabular}{lc|cc|cc|cc|cc} % four columns, alignment for each
\hline
 Model 	&	 Parameters	&	\multicolumn{4}{c}{M1}		&	\multicolumn{4}{c}{M2}	\\
  \cmidrule(lr){3-6} \cmidrule(lr){7-10}																			
	&		&	 Obs 1	&	 Obs 2	&	 Obs 3 	&	 Obs 4  	&	 Obs 1$^c$ 	&	 Obs 2 	&	 Obs 3 	&	 Obs 4 	\\
\hline																			
    \texttt{tbabs} 	&	 $N_H$ ($10^{22}$ cm$^{-2}$) 	&	 $1.1\pm 0.2$ 	&	 $0.54 \pm 0.10$ 	&	 $1.34 \pm 0.15$ 	&	 $1.7 \pm 0.2$ 	&	 $1.08 \pm 0.17$ 	&	 $0.52 \pm 0.11$  	&	 $1.20 \pm 0.17$ 	&	 $1.6 \pm 0.2$ 	\\
																			
    \texttt{Bbodyrad} 1	&	 $kT_{\rm BB}$ (keV) 	&	 $0.32 \pm 0.03$     	&	 $0.366 \pm 0.018$     	&	 $0.281^{+0.015}_{-0.014}$ 	&	 $0.269 \pm 0.015$    	&	 $0.35 \pm 0.03$ 	&	 $0.38 \pm 0.02$ 	&	 $0.31 \pm 0.02$ 	&	 $0.29 \pm 0.02$ 	\\
                   	&	Norm	&	 $1084^{+694}_{-500}$ 	&	 $5705^{+1982}_{-1454}$ 	&	 $22335^{+10531}_{-7119}$ 	&	 $20847^{+12056}_{-7396}$ 	&	 $975^{+682}_{-414}$ 	&	 $5554^{+2112}_{-1484}$ 	&	 $13258^{+7054}_{-4536}$ 	&	 $12465^{+8939}_{-4759}$ 	\\
	&	 $R_{\rm BB}$ (km) 	&	$8.0^{+2.6}_{-1.8}$ 	&	$18.4^{+3.2}_{-2.3}$ 	&	$36.5^{+8.6}_{-5.8}$ 	&	$35.2^{+10.2}_{-6.2}$ 	&	 $7.6_{-1.6}^{+2.7}$ 	&	 $18.2_{-2.4}^{+3.4}$ 	&	 $28.1_{-4.8}^{+7.5}$ 	&	 $27.2_{-5.2}^{+9.8}$ 	\\
																			
    \texttt{Bbodyrad} 2	&	 $kT_{\rm BB}$ (keV) 	&	 $1.58\pm 0.05$     	&	 $1.65\pm 0.03$     	&	 $1.71 \pm 0.03$ 	&	 $1.65\pm 0.03$    	&		&		&		&		\\
                   	&	Norm	&	 $20.9 \pm 2.0$ 	&	 $102.6 \pm 6$ 	&	 $75 \pm 4$ 	&	 $56.5 \pm 3.6$ 	&		&		&		&		\\
	&	 $R_{\rm BB}$ (km) 	&	$1.11 \pm 0.05$ 	&	$2.46 \pm 0.07$ 	&	$2.11 \pm 0.06$ 	&	$1.83 \pm 0.06$ 	&		&		&		&		\\
																			
    \texttt{Gaussian} 	&	 $E_{\rm Line}$ (keV) 	&		&	 $6.33 \pm 0.08$ 	&	 $6.4 \pm 0.1$ 	&	 $6.46^{+0.05}_{-0.08}$ 	&		&	 $6.31 \pm 0.08$ 	&	 $6.39 \pm 0.10$ 	&	 $6.45 \pm 0.06$ 	\\
                   	&	$\sigma$ (keV)	&		&	 $0.15^{+0.08}_{-0.06}$ 	&	 $0.14^{+0.10}_{-0.08}$ 	&	 $0.03^{+0.11}_{-0.03}$ 	&		&	 $0.15^{+0.08}_{-0.05}$ 	&	 $0.13^{+0.09}_{-0.07}$ 	&	 $0.04^{+0.12}_{-0.04}$ 	\\
	&	Norm	&		&	 $5.4_{-1.8}^{+2.0}$ 	&	 $3.5 \pm 1.5$ 	&	 $1.6_{-0.8}^{+1.0}$ 	&		&	 $5.43^{+2.2}_{-1.4}$ 	&	 $3.4^{+1.6}_{-1.1}$ 	&	 $1.75 \pm 0.08$ 	\\
	&	EQW (eV)	&	 $<15$ 	&	 $39 \pm 14$ 	&	 $32 \pm 14$ 	&	 $23 \pm 12$ 	&	$<16$	&	 $39_{-18}^{+14}$ 	&	 $31_{-19}^{+13}$ 	&	 $25_{-13}^{+8}$ 	\\
																			
    \texttt{compTT} 1 	&	 $kT_0$ (keV) 	&	 $3.59 \pm 0.25$ 	&	 $4.04 \pm 0.11$ 	&	 $3.9 \pm 0.1$ 	&	 $3.63 \pm 0.12$ 	&	 $1.22 \pm 0.04$ 	&	 $1.25 \pm 0.04$  	&	 $1.31_{-0.04}^{+0.05}$ 	&	 $1.27 \pm 0.04$ 	\\
                    	&	 $kT_e$ (keV) 	&	 $>12$	&	 $14.1 \pm 0.3$	&	 $17.0 \pm 0.4$	&	 $19.9_{-0.8}^{+0.9}$	&	 $13.2_{-2.1}^{+4.1}$ 	&	 $13.5 \pm 0.3$	&	 $16.0_{-1.0}^{+0.5}$	&	 $6.7_{-1.9}^{+0.6}$	\\
                   	&	 $\tau$ 	&	 $<2$ 	&	 $2.37 \pm 0.09$ 	&	 $2.05 \pm 0.08$ 	&	 $1.74 \pm 0.11$ 	&	 $2.46_{-0.46}^{+0.33}$ 	&	 $3.11_{-0.10}^{+0.15}$  	&	 $2.7_{-0.2}^{+1.7}$ 	&	 $4.3_{-0.2}^{+2.9}$ 	\\
                   	&	Norm ($10^{-3}$)	&	 $4.2_{-0.8}^{+10.1} \times 10^{-4}$	&	 $0.122 \pm 0.003$ 	&	 $0.0696 \pm 0.0025$ 	&	 $0.0334 \pm 0.0019$ 	&	 $0.026_{-0.006}^{+0.005}$  	&	 $0.174_{-0.001}^{+0.004}$  	&	 $0.102_{-0.04}^{+0.02}$ 	&	 $0.17 \pm 0.05$ 	\\
																			
    \texttt{compTT} 2 	&	 $kT_e$ (keV) 	&		&		&		&		&	 - 	&	 $5.5_{-0.25}^{+0.04}$ 	&	$5.8_{-0.7}^{+1.1}$ 	&	 $15.8_{-0.9}^{+3.0}$ 	\\
                   	&	 $\tau$ 	&		&		&		&		&	 - 	&	 $200^{p}$  	&	$8.2_{-3.6}^{p}$ 	&	 $6.0_{-3.9}^{p}$ 	\\
                   	&	 Norm 	&		&		&		&		&	 -  	&	 $0.054_{-0.004}^{+0.008}$  	&	 $0.068_{-0.04}^{+0.18}$ 	&	 $0.013_{-0.008}^{+0.047}$ 	\\
																			
    \texttt{edge} 	&	 $E$ (keV) 	&		&		&		&		&	 $9.62 \pm 0.72$ 	&	 $9.7 \pm 0.4$ 	&	 $9.9 \pm 0.4$ 	&	 $9.9 \pm 0.4$ 	\\
                   	&	 depth 	&		&		&		&		&	 $0.10 \pm 0.04$ 	&	 $0.09 \pm 0.03$	&	 $0.09 \pm 0.03$  	&	 $0.09 \pm 0.02$  	\\
																			
                   	&	 $C_{\rm SXT}$ 	&	 $1.12 \pm 0.03$ 	&	 $0.83 \pm 0.01$ 	&	 $1.07 \pm 0.01$ 	&	 $1.09 \pm 0.01$	&	 $1.12 \pm 0.03$ 	&	 $0.83 \pm 0.01$ 	&	 $1.08 \pm 0.01$ 	&	 $1.10 \pm 0.01$ 	\\
                   	&	 $C_{\rm CZTI}$ 	&	 $1.08 \pm 0.08$               	&	 $1.079 \pm 0.015$               	&	 $1.026 \pm 0.014$ 	&	 $1.06 \pm 0.02$           	&	 $1.03 \pm 0.07$ 	&	 $1.08 \pm 0.02$ 	&	 $1.03 \pm 0.02$ 	&	 $1.06 \pm 0.02$ 	\\
%	&	 Gain offset (eV) 	&	 $11.6$	&	 $18_{-23}^{+22}$	&	 $-27.6 \pm 30$ 	&	 $-51.6_{-29}^{+18}$	&	 $11$ 	&	 $20$ 	&	 $-31_{-40}^{+34}$ 	&	 $-55_{-26}^{+29}$ 	\\
                   	&	 Flux$_{\rm 1-150~keV}^a$ 	&	 $4.7 \times 10^{-9}$	&	 $4.7 \times 10^{-8}$	&	 $3.6 \times 10^{-8}$ 	&	 $2.1 \times 10^{-8}$ 	&	 $4.5 \times 10^{-9}$ 	&	 $4.7 \times 10^{-8}$ 	&	 $3.6 \times 10^{-8}$ 	&	 $2.1 \times 10^{-8}$ 	\\
                   	&	 Luminosity$_{\rm 1-150~keV}^b$ 	&	 $3.3 \times 10^{36}$ 	&	 $3.3 \times 10^{37}$ 	&	 $2.6 \times 10^{37}$ 	&	 $1.5 \times 10^{37}$ 	&	 $3.2 \times 10^{36}$ 	&	 $3.3 \times 10^{37}$ 	&	 $2.6 \times 10^{37}$ 	&	 $1.5 \times 10^{37}$ 	\\
\hline                   																			
                   	&	 $\chi^2$/dof 	&	141.5/110	&	146.9/129             	&	155.5/127 	&	172.9/126                 	&	142.7/110	&	137.6/126	&	137.5/123	&	153.2/123	\\
\hline																			
																		
		\multicolumn{10}{l}{$^p$pegged.}\\
		\multicolumn{10}{l}{$^a$Unabsorbed flux in the units of \erg.}\\
    \multicolumn{10}{l}{$^b$Unabsorbed X-ray luminosity in the units of \lum, assuming a source distance of 2.44 kpc.}\\
      \multicolumn{10}{l}{$^c$Only one \texttt{compTT} component was required.}\\
	\end{tabular}}
\end{table*}

\begin{figure*}
\centering
 \includegraphics[width=0.26\linewidth]{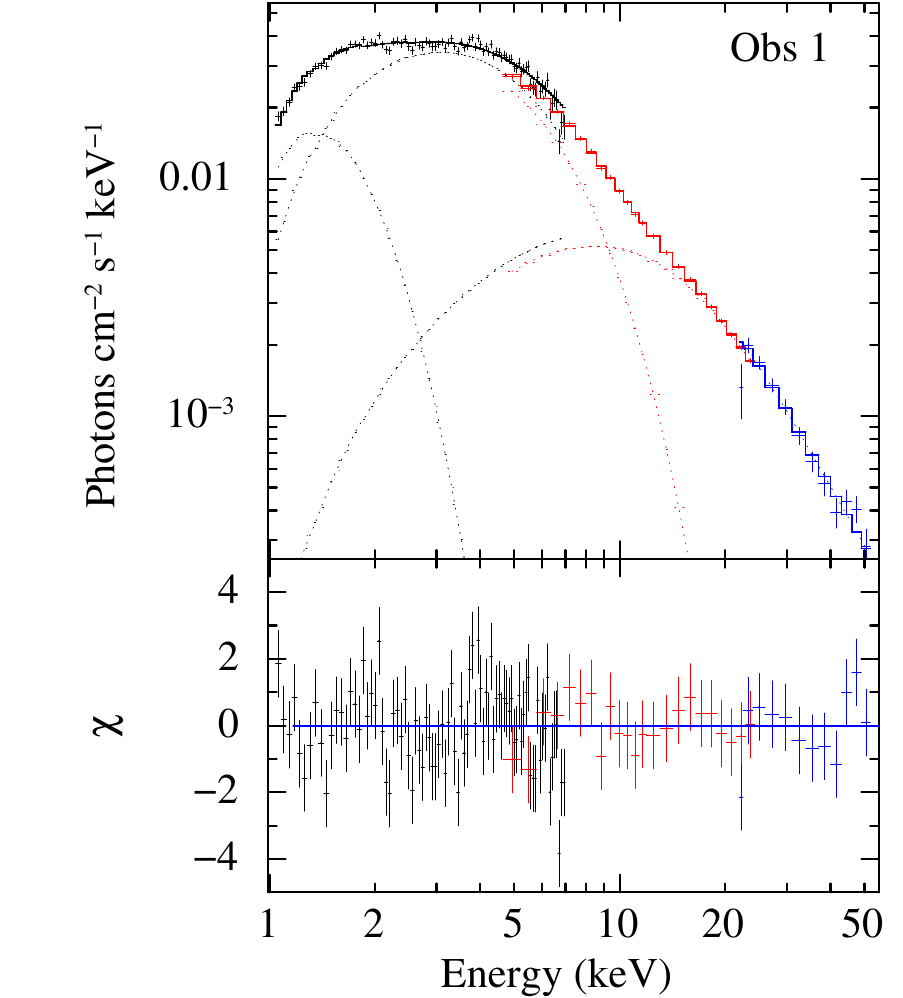}
  \includegraphics[width=0.23\linewidth]{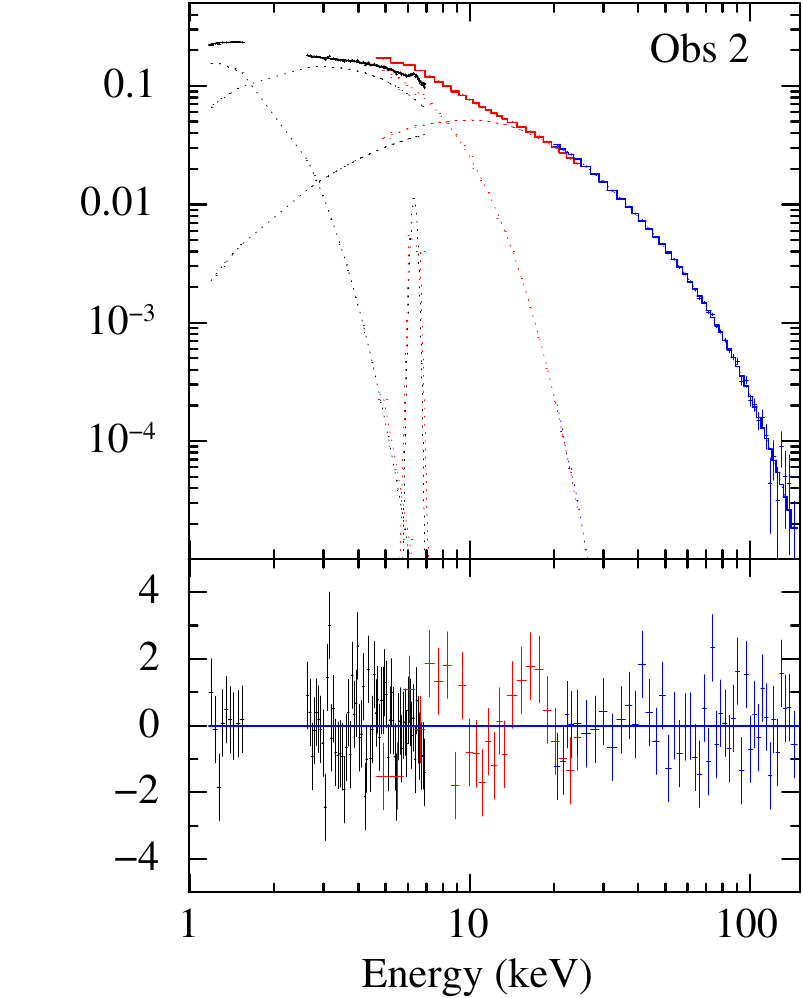}
   \includegraphics[width=0.23\linewidth]{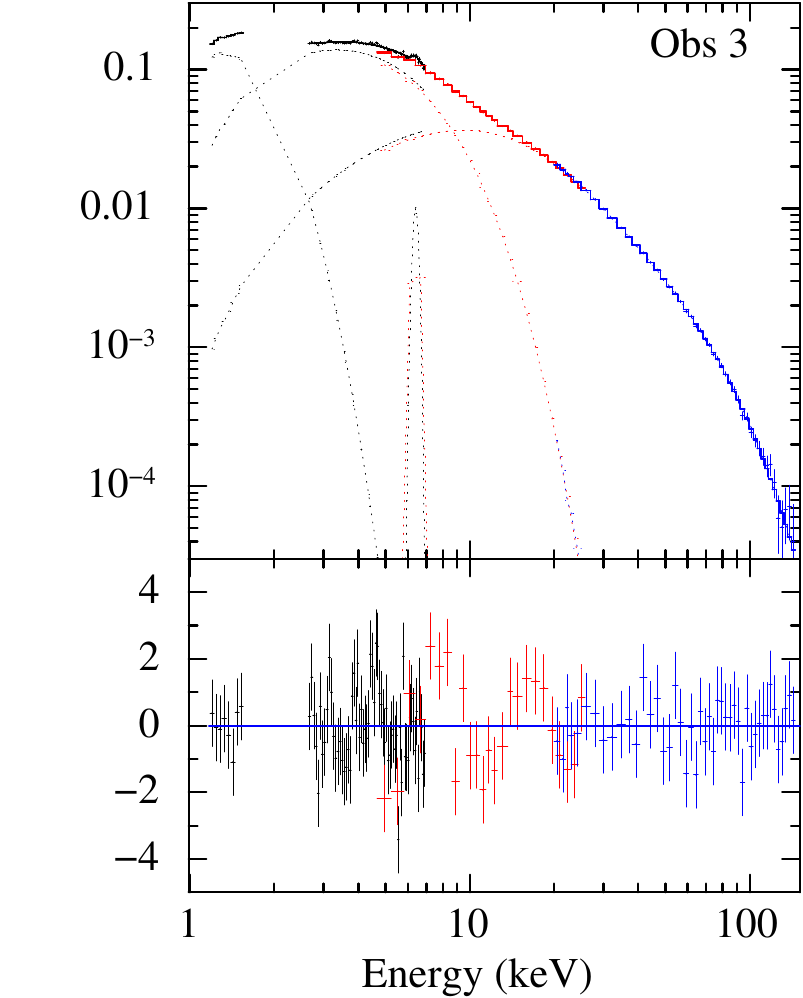}
    \includegraphics[width=0.23\linewidth]{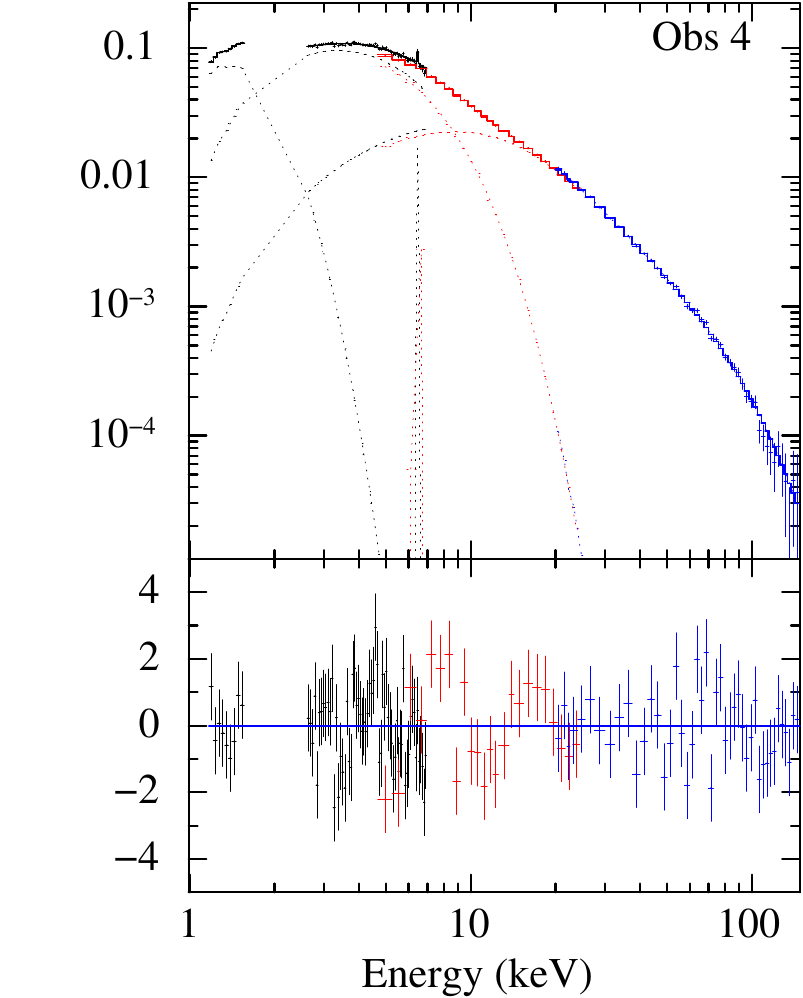}
 \caption{Best-fit broadband energy spectra using SXT, LAXPC, and CZTI fitted with the model M1 (\texttt{tbabs$\times$(bbodyrad+bbodyrad+compTT+Gaussian)}).}
\label{fig:spec}
\end{figure*}

\subsection{Broad-band spectral analysis}
\label{spectral}

To study the broadband spectrum of \src{}, we performed a combined spectral analysis of SXT, LAXPC20, and CZTI data. The spectra were modelled and analyzed using {\tt XSPEC} version 12.13.0c \citep{Arnaud1996}. We consider the LAXPC20 data up to 25 keV for the spectral fitting due to larger uncertainty in background estimation around the K-fluorescence energy of Xe at 30 keV \citep{Antia2017, Sharma2020, Beri2021}. The SXT data in the energy range of 1--7 keV are used for the combined spectral fitting. The CZTI spectra are used for 20--150 keV for obs. 2 to 4, and for observation 1, the CZTI spectra are used up to 50 keV due to poor source count statistics above 50 keV. The SXT, LAXPC, and CZTI spectra were grouped using \textsc{ftgrouppha} using the optimal binning scheme of \citet{Kaastra2016} with a minimum of 25 counts per bin. We added a constant component to represent the cross-calibration between the three instruments. During the fitting, the constant was fixed to 1 for the LAXPC and allowed it to vary for SXT and CZTI. 
After the launch of \astrosat, the gain of the SXT instrument has been changed by a few tens of eV \citep[e.g.,][]{Beri2021, Beri2023}, and as per suggestions of the SXT team, we applied a gain correction to the SXT spectra with a slope fixed to 1.0 and a best-fitting offset between -50 eV to 20 eV. A systematic uncertainty of 1--2\% was used during spectral fitting. The multiplicative model \texttt{tbabs} was used to account for the line-of-sight absorption with abundance from \citet{Wilm2000}.

The X-ray spectra of X-ray pulsars are usually described by the cutoff power law (\texttt{cutoffpl}) or power law with high-energy exponential cutoff (\texttt{highecut*powerlaw}), a combination of two negative and positive power laws with exponential cutoff (\texttt{NPEX}), and thermal Compotnization component (\texttt{compTT}). The broadband spectra from \astrosat{} could not be adequately fitted with a single continuum model. Although adding a soft thermal component improved the fit statistics, it was still not satisfactory, except for observation 1. However, the addition of another soft component provided a satisfactory fit to the spectra. Irrespective of the model used, SXT spectra of Obs 2, 3, and 4 showed emission-like features around 1.9 keV and 2.3 keV, possibly due to mismatch calibration of Si K and Au M edges. Consequently, we excluded the 1.6--2.6 keV energy range from the SXT spectra of observations 2, 3, and 4. For Obs 1, which had the lowest flux, such features were not present. We also ignore below 1.15 keV of SXT spectra to mitigate low-energy calibration issues.

The broadband spectra from SXT+LAXPC+CZTI can be well described with a \texttt{bbodyrad+bbodyrad+compTT} model (hereafter, Model M1). The thermally Comptonized model \texttt{compTT} \citep{Ti94} is characterized by the soft seed photons at temperature $kT_0$, which are Comptonized with hot plasma at temperature $kT_e$ and optical depth $\tau$. Following \citet{Li23, Sa23}, we also tested a two-component \texttt{compTT} model with a soft thermal component model (Model M2). Model M2 required an additional edge component at 10 keV to fully describe the continuum and provide a satisfactory fit \citep{Manikantan2023}. For Obs 1, a second \texttt{compTT} component was not necessary. Additionally, a neutral Fe K emission line was detected in Obs 2, 3, and 4 with equivalent widths of $\sim$23--40 eV. For Obs1, an upper limit on the Fe-line equivalent width of $<$16 eV was found.

The best-fit broadband energy spectra for four different \astrosat{} observations at different phases of the outburst are shown in Fig. \ref{fig:spec}. Table \ref{tab:fitstat2} summarizes the best-fit spectral parameters for models M1 and M2. For Model M1, the soft thermal blackbody temperature ranged between $\sim$0.27--0.37 keV, while the hot blackbody temperature was $\sim$1.6--1.7 keV. The radius of the blackbody emission region was estimated from the normalization of the blackbody component, which is scaled by $(R_{\rm BB}/D_{10})^2$, where $D_{10}$ is the distance in units of 10 kpc (see Table \ref{tab:fitstat2}).
The \texttt{compTT} component suggested a very hot seed source with a temperature of $\sim$3.6--4.0 keV, an electron temperature ($kT_e$) of 12--20 keV, and a plasma optical depth of $\sim2$. However, for Obs 1, the electron temperature and plasma optical depth were not well-constrained. In Model M2, the thermal component was similarly soft, with temperatures akin to those in Model M1. The seed photon temperatures ($kT_0$) were tied together for both \texttt{compTT} humps and ranged between 1.2--1.3 keV. The \texttt{compTT} components exhibited a plasma temperature of $\sim$6--7 keV and $\sim$13--16 keV. The unabsorbed X-ray luminosity in the energy range of 1--150 keV varied between (0.33--3.3)$\times10^{37}$ erg s$^{-1}$ during the four \astrosat{} observations, for a source distance of 2.44 kpc \citep{Ba21}.

%%%%%%%%%%%%%%%%%%%%%%%%%%%%%%%%%%%%%%%%%%%%%%%%%%%%%%%%%%%%%%%%%%%%%%%%%%%%%%%%
%%%%%%%%%%%%%%%%%%%%%%%%%%%%%%%%%%%%%%%%%%%%%%%%%%%%%%%%%%%%%%%%%%%%%%%%%%%%%%%%
%%%%%%%%%%%%%%%%%%%%%%%%%%%%%%%%%%%%%%%%%%%%%%%%%%%%%%%%%%%%%%%%%%%%%%%%%%%%%%%%

\section{Discussion and Conclusions}
\label{dis}

We report the findings of spectral and temporal analysis of \src{} using data from \astrosat{} observations during its giant outburst of 2022--2023. A state transition from the subcritical to the supercritical accretion regime was identified, and the source exhibited a significant evolution in the hardness ratio \citep{Mandal23, Li23, Sa23}.  
The spectral-timing results and source luminosity measured during the \astrosat{} observations indicate that Obs 1, 3, and 4 were in the subcritical regime, while Obs 2 was in the supercritical regime. These \astrosat{} observations allowed us to probe the broadband nature of the source at different accretion regimes.

\subsection{Quasi-periodic variability}
\label{sec4.1}
\begin{table}
	\centering
	\caption{Characteristics of the observed QPOs.}
	\label{tab:mf}
	\resizebox{0.65\columnwidth}{!}{
	\begin{tabular}{lcc} % four columns, alignment for each
		\hline
	 QPO & Obs 3 & Obs 4   \\
		\hline
    $\nu_k$ (mHz)&  60 & 42\\
    $r_k$ (km)& $10.9 \times 10^3$ & $13.9 \times 10^3$\\
    $L_X$ (\lum) & $2.6 \times 10^{37}$ & $1.5 \times 10^{37}$\\
    $\dot{m}$ (g s$^{-1}$) & $1.4 \times 10^{17}$ & $0.8 \times 10^{17}$\\
    $B$ (G) & $7.3 \times 10^{13}$ & $8.4 \times 10^{13}$\\    
		\hline
	\end{tabular}}
\end{table}

QPOs are typically exhibited by accreting high magnetic field X-ray pulsars at frequencies of a few tens of mHz and are associated with phenomena related to the inner accretion disc \citep{Angelini1989, Finger1996, Paul1998, Kaur2007, Devasia2011, Sharma23, Sharma2023b, Manikantan2024}. The characteristics of QPOs and their evolution with time and energy have been used to understand the physical processes leading to QPO generation, the size of the emitting region, and their connection with the spectral parameters \citep{Finger1998, Manikantan2024}. We detected QPOs in \src\ at 60 mHz and 42 mHz when the source was detected at a luminosity of $2.6\times 10^{37}$ \lum\ and $1.5 \times 10^{37}$ \lum, respectively (Table \ref{tab:mf}). Both QPOs were found in the subcritical phase (Obs 3 and 4). Previously, QPOs have been detected in the subcritical regimes of some source, such as in KS 1947+300 \citep{James2010}, while in some other X-ray pulsars, QPOs were detected in supercritical regime e.g., 1A 0535+262 \citep{Finger1996, Reig2008} and EXO 2030+375 \citep{Angelini1989}. The non-detection of QPOs in the first two observations suggests that QPO presence is a transient phenomenon, similar to several other accreting pulsars %such as Cen X-3, KS 1947+300, 1A 1118-615 and LMC X-4 
\citep{Raichur2008, James2010, Nespoli2011, Rikame2022, Sharma23, Sharma2023b, Chhotaray2024, Manikantan2024}.

The observed QPOs in X-ray pulsars can be explained with the most commonly used models: the Keplerian frequency model \citep[KFM;][]{vanderKlis1989} in which QPOs result from an accreting inhomogeneity in the accretion disc rotating at the magnetospheric radius with the Keplerian disc frequency, $\nu_{\rm QPO} = \nu_k$; and the beat frequency model \citep[BFM;][]{Alpar1985, Lamb1985}, which attributes QPOs to the modulations in the mass accretion rate at the beat frequency between the orbital frequency of the inner accretion disc and spin frequency of the neutron star, $\nu_{\rm QPO} = \nu_{\rm beat} = \nu_k - \nu_{\rm spin}$. As the radius of the inner accretion disc scales by the mass accretion rate or the X-ray luminosity, the Keplerian frequency of the inner accretion disc (and thereby the QPO frequency) is expected to show a positive correlation with luminosity in both KFM and BFM \citep{Angelini1989, Finger1996, Finger1998, Mukherjee2006}.
  
\citet{Malacaria24} and \citet{Li2024} reported pulse phase-dependent transient QPOs in \src{} at $\sim$0.2 Hz and 0.14--0.5 Hz, respectively, when source luminosity (1--79 keV) was $\gtrsim3\times 10^{37}$ \lum.  
Above this luminosity, during the right wing of pulse profile peaks, flares were generated on a timescale of seconds by an increase in the hard photons and have been detected with \textit{Fermi}-GBM and \textit{Insight}-HXMT \citep{Malacaria24, Li2024}. The 0.2--0.5 Hz QPOs were manifested as large modulations in these flares. These QPOs, however, did not depend on the source luminosity and cannot be explained with the KFM or BFM \citep{Li2024}. The QPOs detected in the current work show luminosity dependence, where QPO frequency decreases from 60 mHz to 42 mHz with a decrease in luminosity. Therefore, both low-frequency QPOs can be explained with the KFM or BFM. 
The non-detection of mHz QPOs in previous studies may be linked to the evolution of physical conditions within the accretion disc, particularly changes in viscosity and the propagation of instabilities. Specifically, the hot part of the disc effectively contributes to the propagating fluctuations in the mass accretion rate. The evolution of the outburst and dynamics of the accretion disc can affect the PDS of aperiodic variability and the occurrence of QPOs at low frequencies \citep{Mushtukov2019}. 

The Keplerian frequency at the inner disc radius $r_k$ is given by
\begin{eqnarray}
    r_k = \Big( \frac{G~M_{\rm NS}}{4~\pi^2~\nu_k^2} \Big)^{1/3} 
\end{eqnarray}
where $M_{\rm NS}$ is the neutron star mass. The inner accretion disc is expected to terminate at the magnetospheric radius ($r_m$) or Alfven radius ($r_A = r_m/k$), where the energy density of the magnetic field balances the kinetic energy density of the infalling material \citep{Davies1981}. Since the QPO frequency is much higher than the spin frequency, $\nu_{\rm QPO} \sim \nu_k$, KFM and BFM are indistinguishable and have similar radii at which the QPOs are generated. Therefore, assuming $r_k=r_m$, the magnetic field strength ($B = 2\mu/R^3$) at the surface of a neutron star can be estimated using the formula,
\begin{eqnarray}
    B = 2.31 \times 10^{12} ~\Big(\frac{k}{0.5}\Big)^{-7/4} ~\Big(\frac{M_{\rm NS}}{1.4 M_\odot}\Big)^{5/6} ~R_6^{-3} ~\dot{m}_{17}^{1/2} ~\nu_k^{-7/6} ~{\rm G} 
\end{eqnarray}
where $k$ is the coupling constant usually taken to be 0.5 for disc accretion \citep{Ghosh1979, Mushtukov2022}, $R_6$ is the radius of the neutron star in units of $10^6$ cm, and $\dot{m}_{17}$ is the mass accretion rate in units of $10^{17}$ g \psec. The X-ray luminosity in the 1--150 keV was used to estimate the mass accretion rate ($\dot{m}$) using $L_X = G M_{\rm NS} \dot{m}/R = \eta \dot{m} c^2$, where $\eta$ is the accretion efficiency, assumed to be 0.207 \citep{Sibgatullin2000} for a neutron star with a radius of 10 km and mass of 1.4 $M_\odot$. The estimated value of the inner accretion disc radius and the magnetic field strength for the two observed QPOs are given in Table \ref{tab:mf}. \src\ is inferred to have a relatively high magnetic field of the order of $10^{13}$ G compared to typical X-ray pulsars. However, the estimated magnetic field strength is subject to uncertainties from various variables such as coupling constant $k$ and accretion efficiency $\eta$. Previous studies also suggested a comparatively higher magnetic field strength ($\sim10^{13}$ G) using different approaches for this X-ray pulsar \citep{Li2024, Malacaria24, Mandal23, Sa23}.

\subsubsection{Energy-dependence of QPO}

Both QPOs showed energy dependence in the broad energy band of 3--80 keV (Fig. \ref{fig:e-qpo}). The rms of both QPOs first decreased with energy up to 10--15 keV, then increased thereafter to 25\% and 15\% for 60 mHz and 42 mHz QPOs, respectively. Some X-ray pulsars, e.g., KS 1947+300, XTE J1858+034, IGR J19294+1816, 4U 1626--67, have shown a correlation between QPO rms and energy \citep[e.g.,][]{James2010, Mukherjee2006, Manikantan2024}. However, such behaviour is not seen in general.
The increasing trend of rms with energy in \src{} above 15 keV indicates that these aperiodic oscillations are more likely due to variations in the hot plasma \citep{Sharma23}.

Previously, no trends in the QPO parameters, such as centre frequency and width, except the fractional rms amplitude, as a function of energy in X-ray pulsars were found \citep[e.g.,][]{Finger1998, James2010, Ma2022, Sharma23, Manikantan2024}. The mHz QPO observed in \src{} clearly showed energy dependence on the QPO frequency as well. 
The QPO frequency for the 60 mHz QPO varied between 56 mHz and 61 mHz, showing an increasing trend up to 10--15 keV, followed by a decrease. Meanwhile, for QPO at 42 mHz, a decreasing trend in the QPO frequency was observed with energy. The QPO frequency decreased from $43.6 (6)$ mHz at 3--5 keV to $39.2 (7)$ mHz above 25 keV. Additionally, an increase in rms is observed with the decrease in the QPO frequency (see Fig. \ref{apx:freq-rms}).
These properties pose a challenge for the KFM and BFM. More sophisticated models are required to elucidate an explanation of the mHz QPO observed \citep{Bozzo2009, Ma2022}.

\subsection{Pulse timing properties}
During the outburst, \astrosat{} observations were used to study the evolution of the spin period, pulse profile, rms PF, and emission mechanism of \src. We found that the pulse profile evolves significantly with energy and luminosity. Near the peak of the outburst (above critical luminosity), the pulse profile evolves from a dual-peak feature to a broad single-peak feature consisting of a minor peak. A multi-peak pattern with strong energy dependence re-emerges in the pulse profile during the decay phase of the outburst. 
A similar behaviour has been observed in other X-ray pulsars \citep[e.g.,][]{Beri2017, Epili2017, Wilson2018, Chhotaray2024}.

A phase difference was observed between the high-energy peak and low-energy peak of the pulse profiles, varying with luminosity. The difference in peak phase is more prominent during outburst decay (Obs 3 and 4), suggesting that high-energy photons and low-energy photons do not originate from the same direction. Earlier, several studies were performed using \nicer{}, \insighthxmt{}, \nustar{} to probe the evolution of the pulse profile, and it was found that the pulse profile evolved from a double peak feature to a single peak feature above the critical luminosity \citep{Mandal23, Li23, Sa23}, which is consistent with our results.   
The pulse profiles are found to depend on both the energy and luminosity, probably indicating a change in accretion mode as well as in the beaming patterns of the source and their dependence on the energy and mass accretion rate. The changes in the emission pattern typically lead to variations in the observed pulse profile shape and the spectral shape \citep{Mandal23}.
Additionally, an absorption dip is evident in the pulse profiles during the subcritical regime, which diminishes with increasing energy and disappears in hard X-rays. This dip is most prominent in the SXT light curves (Fig. \ref{fig:pp-all}), likely caused by the absorption or scattering of X-ray emission by the accretion stream above the neutron star \citep{Ts12, Usui2012, Sa23}. However, this absorption dip is not observed in the supercritical regime, probably due to a change in the geometry of the accretion stream with critical luminosity. 
    
The rms PF also shows a significant variation in energy and luminosity. The PF decreases with energy up to $\sim$25 keV, and above this energy, PF shows an increasing trend with energy, similar to other X-ray pulsars \citep{Lutovinov2009}. A local minimum feature in the variation of PF with energy is evident around the fluorescence iron line \citep{Sa23}. Earlier, the detailed timing study with {\it Insight}/HXMT showed a concave-like feature around 20--30 keV in the plot of the PF \citep{Li23}. The concave-like feature was more prominent below the critical luminosity. The concave was weaker above the critical luminosity, which is probably due to the lower fraction of reflected photons and the higher accretion column \citep{Li23}. 
The \astrosat{} timing results also showed this concave-like feature around 25 keV. Above the critical luminosity (supercritical regime), the accretion column is higher than the lower luminosity (subcritical regime), and the accretion column results in a lower probability of photon scattering, which results in a weaker concave. The concave near 20--30 keV suggests a dilution of reflected photons over this energy range. The concave depth is linked with the proportion of reflected photons, which is also related to the height of the accretion column \citep{Poutanen2013}. 

\subsection{Spectral properties}

The broadband spectra of \src{} were studied using SXT, LAXPC, and CZTI during four different observations at various phases of the outburst. During Obs 1, the source was detected up to 50 keV, while for the subsequent observations, the detection extended up to 150 keV. The broadband continuum of \src{} can be well described using two thermal components with a Componized emission model (M1) or a two-component Comptonization model (\texttt{compTT+compTT}) with a soft thermal component (M2). The \texttt{compTT} spectra in accreting X-ray pulsars are generally considered to be a consequence of thermal Comptonization processes, wherein the seed photons originating from the NS hotspots get the thermal energy of the accreting gas \citep{Becker2007}. 

In Model M1, two thermal blackbody components yielded temperatures of $\sim$0.3 and $\sim$1.6 keV with emission radii of tens of km and 1--2 km, respectively. These two blackbody components may be associated with the thermal emission from the photosphere of optically thick outflows and from the top of the accretion column, respectively. The \texttt{compTT} component had a seed source temperature of $\sim$3.6--4 keV with a plasma temperature of $\sim$12--20 keV. This hot seed emission likely originates from the hotspot of the neutron star \citep[see,][]{Tao2019, Beri2021}. 

In Model M2, a two-component continuum model (\texttt{compTT} + \texttt{compTT}) with a soft thermal component along with an edge was used. The \astrosat{} spectra suggested a seed photon temperature of $\sim$1.3 keV tied for the two humps. The two Comptonization components showed electron temperatures of 13--16 keV for hot and 5--6 keV for relatively colder components. For Obs 1, 2, and 3, the hotter component was dominating, while for Obs 4, the colder component became more significant. A similar model with two components (\texttt{compTT} + \texttt{compTT}) with blackbody has been previously used to describe the broadband spectra of \src{} \citep{Li23, Sa23}. However, this two-component compTT model is generally used to describe the X-ray pulsar spectra at low luminosity ($\lesssim10^{35}$ \lum) \citep{Doroshenko2012, Tsygankov2019, Tsygankov2019b, Lutovinov2021, Sokolova-Lapa2021}, but also see, \citet{Doroshenko2020}. The exact origin of the high-energy component is unclear but is likely due to the combined effect of cyclotron emission and thermally Comptonized emission from a thin, overheated layer of the NS atmosphere \citep{Tsygankov2019}. 

The iron emission line was significantly detected in Obs 2, 3, and 4, irrespective of the model used, with equivalent widths to be low $\sim$23--40 eV. A similar low equivalent width of the Fe line was found in disc-fed X-ray pulsars such as SMC X--1, LMC X--4, and 4U 1626--67 \citep{Paul2002, Sharma2023b}. However, \nustar{} spectra suggest a higher equivalent width for the Fe-line at higher luminosity \citep{Sa23}.

The source was at different accretion luminosity ($L_X \sim 0.33-3.3 \times 10^{37}$ \lum) during \astrosat{} observations, with a state transition between the subcritical and supercritical regime. These two accretion regimes are defined by the critical luminosity, above which the beaming patterns of the pulsar and accretion mechanism evolved significantly \citep{Basko1975, Basko1976, Becker2012, Mushtukov2015}. For \src, the critical luminosity was found to be $\sim3\times 10^{37}$ \lum\ \citep{Mandal23, Li23, Sa23}. During the outburst, the \astrosat{} observations implied a significant change in pulse profiles above a certain luminosity, corresponding to the change in accretion regime and resulting in a possible change in the beaming patterns. Also, we did not find any absorption line due to the cyclotron resonance scattering feature at hard X-rays, coherent with a magnetic field strength greater than $10^{13}$ G \citep{Sa23}. 

\section*{Acknowledgements}
Data from the ToO phase of \astrosat{} observations were used in this study. This research has made use of the \astrosat\ data obtained from the Indian Space Science Data Centre (ISSDC). We thank the LAXPC, CZTI Payload Operation Centers (POCs), and the SXT POC at TIFR, Mumbai, for providing the necessary software tools. 
We also thank the anonymous referee for valuable comments and suggestions on the manuscript.

%%%%%%%%%%%%%%%%%%%%%%%%%%%%%%%%%%%%%%%%%%%%%%%%%%
\section*{Data Availability}

Data used in this work can be accessed through the Indian Space Science Data Center (ISSDC) at 
\url{https://astrobrowse.issdc.gov.in/astro\_archive/archive/Home.jsp}.



\bibliographystyle{mnras}
\bibliography{example} % if your bibtex file is called example.bib

% Alternatively, you could enter them by hand, like this:
%\begin{thebibliography}{99}
%\end{thebibliography}
%%%%%%%%%%%%%%%%%%%%%%%%%%%%%%%%%%%%%%%%%%%%%%%%%%

%%%%%%%%%%%%%%%%% APPENDICES %%%%%%%%%%%%%%%%%%%%%

\appendix

\section{Additional Pulse profiles and QPO plots}

\begin{figure*}
\centering
 \includegraphics[width=0.24\linewidth]{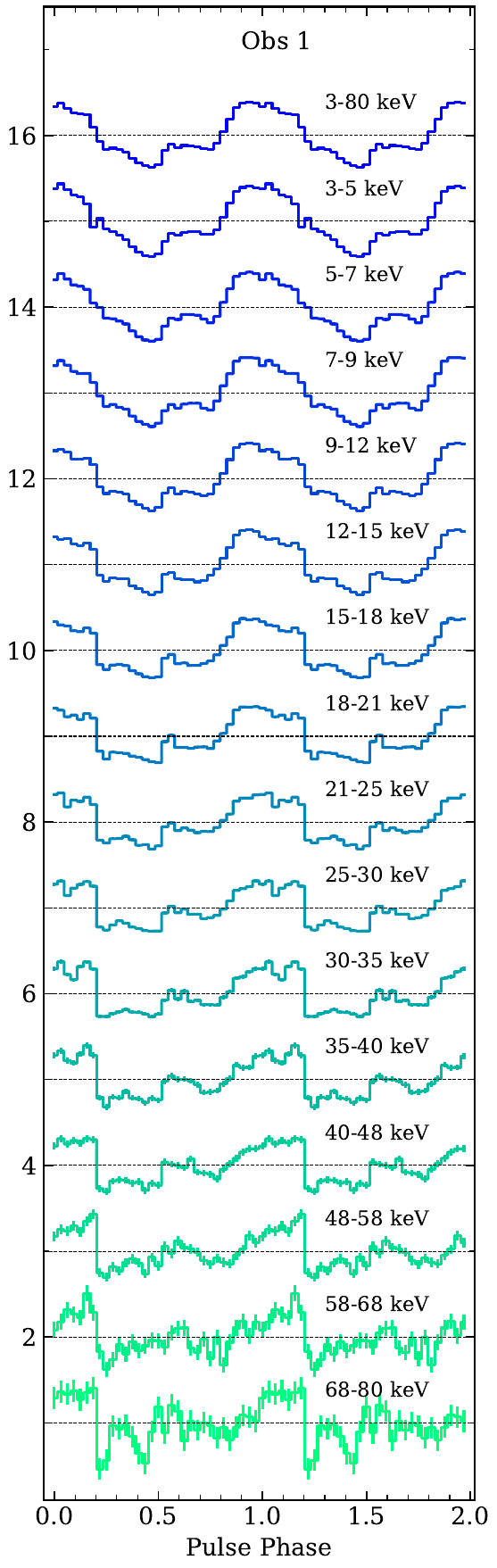}
  \includegraphics[width=0.24\linewidth]{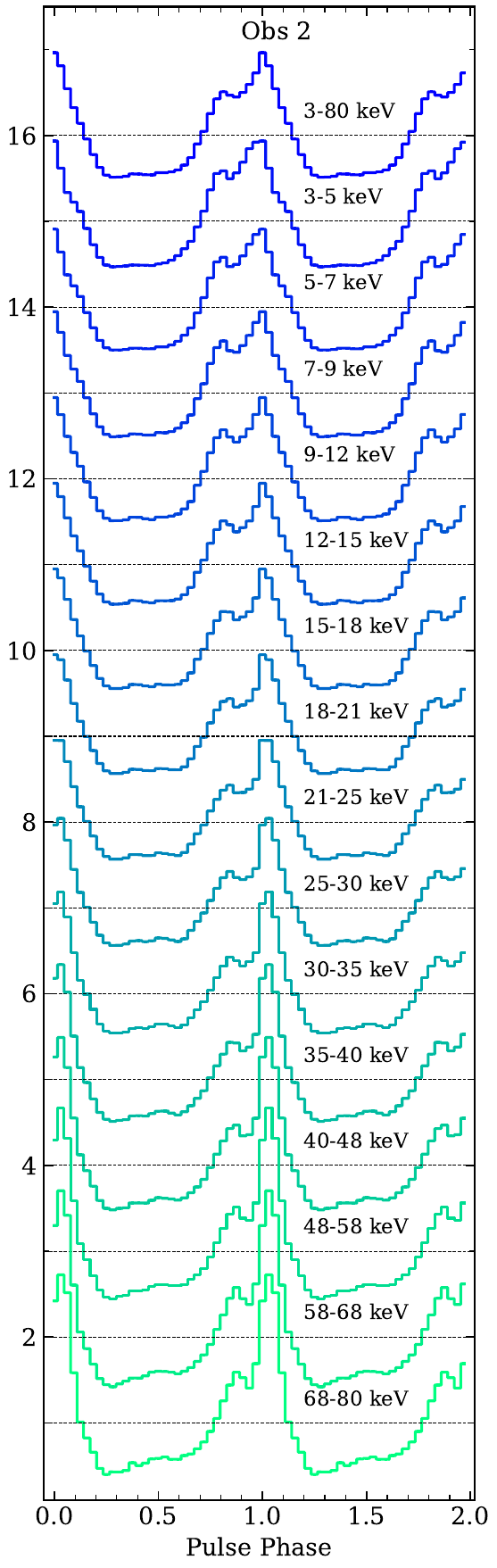}
   \includegraphics[width=0.24\linewidth]{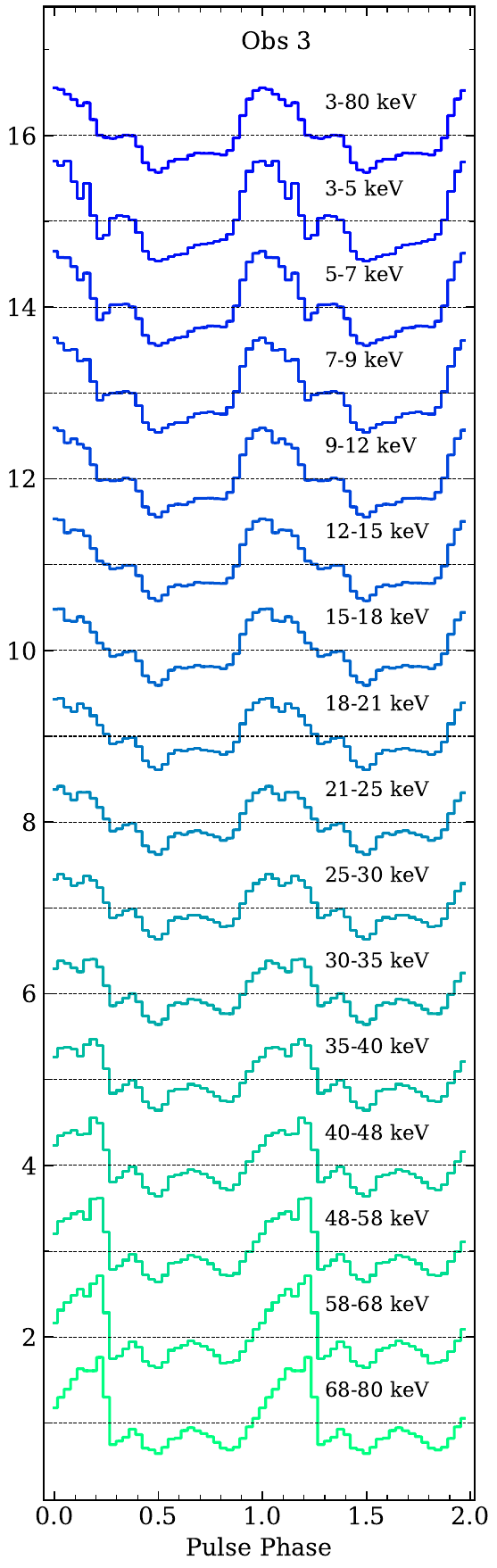}
    \includegraphics[width=0.24\linewidth]{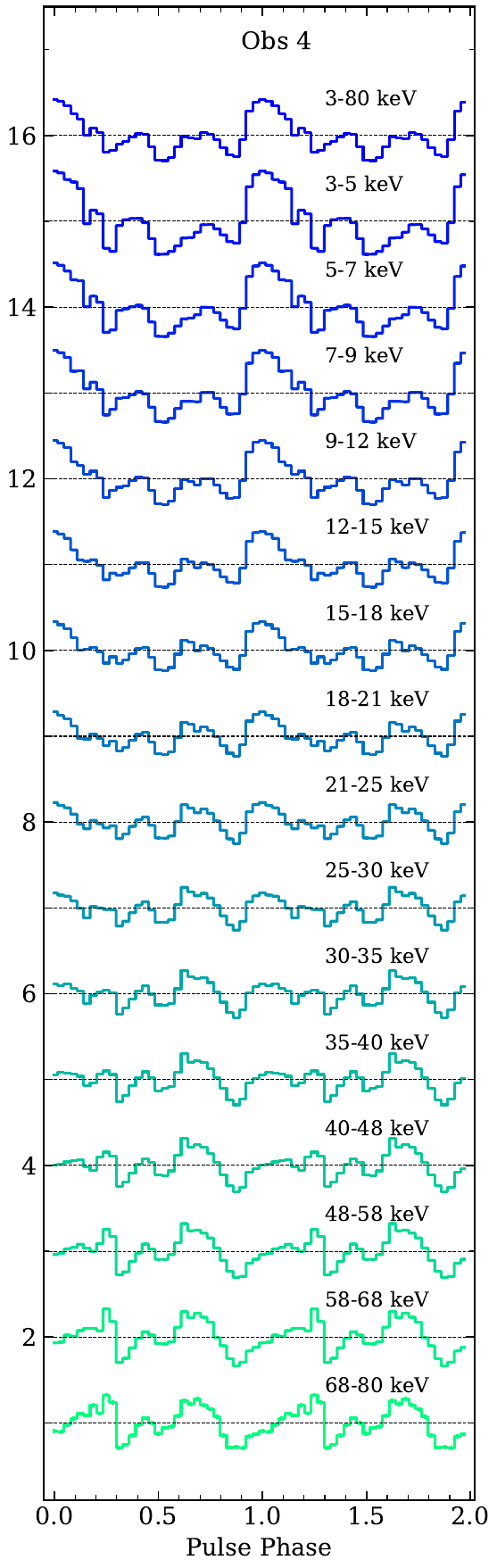}
 \caption{Energy-resolved pulse profiles of \src{} for four different \astrosat/LAXPC observations at different luminosity levels of the outburst. The y-axis represents the normalized intensity of pulse profiles and has been shifted to show each profile clearly. }
\label{apx:pp}
\end{figure*}

\begin{figure*}
\centering{
\includegraphics[width=0.65\columnwidth]{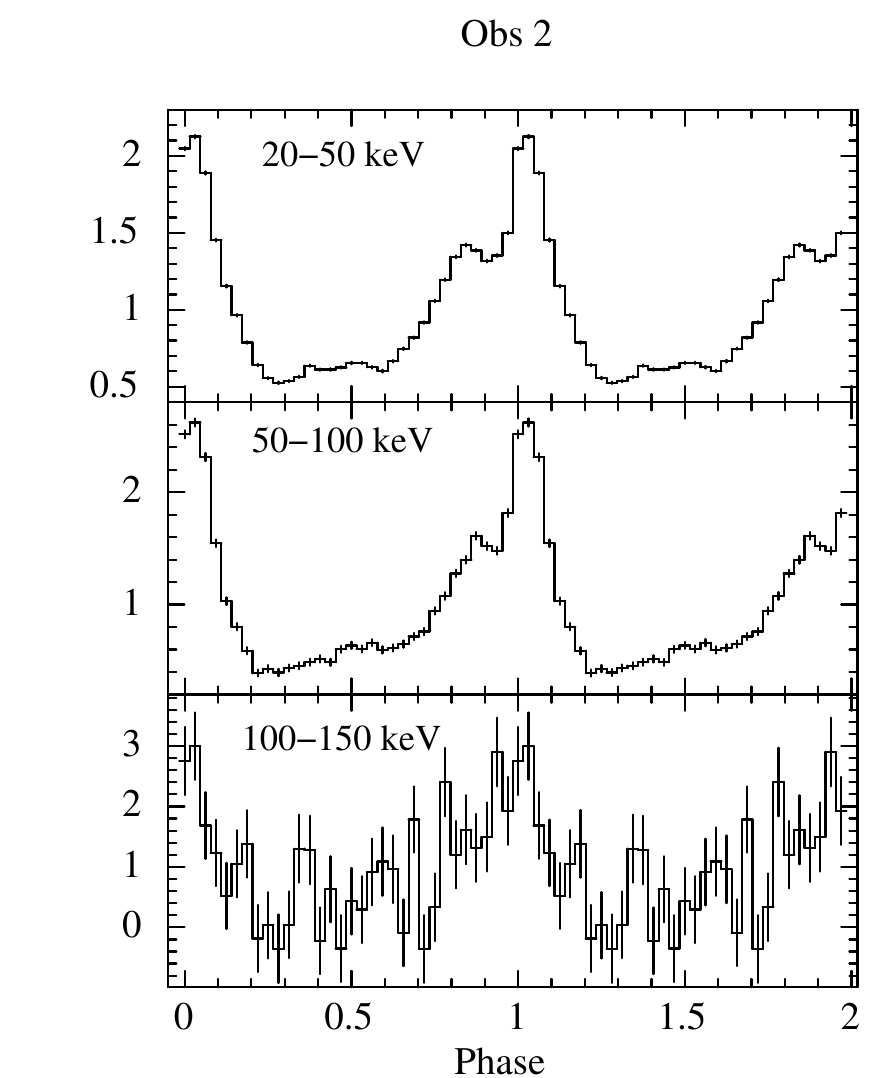}
\includegraphics[width=0.65\columnwidth]{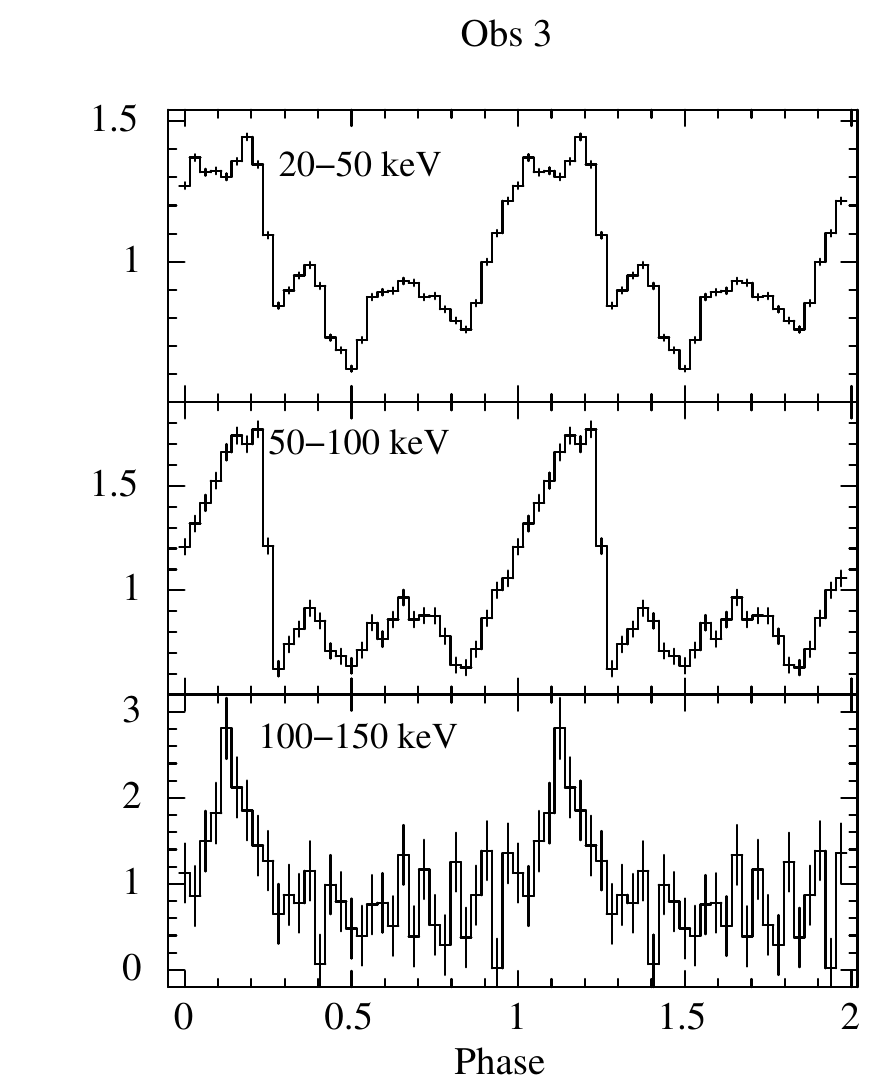}
\includegraphics[width=0.65\columnwidth]{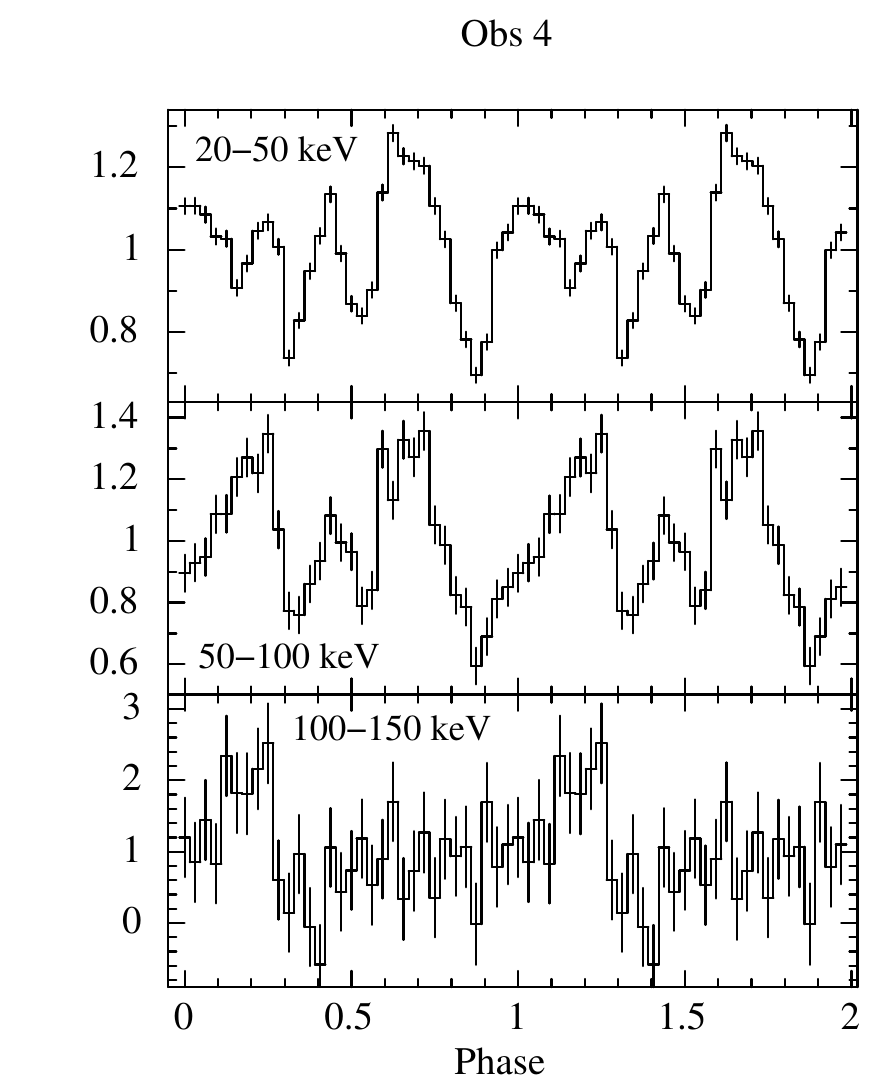}
\caption{Energy resolved pulse profiles of \src{} using \astrosat/CZTI observations during peak and decay phases of the outburst. Pulsations were detected up to 150 keV in observations 2, 3, and 4. The y-axis represents the normalized intensity of pulse profiles.}
    \label{apx:pp_CZTI}}
\end{figure*}

\begin{figure*}
\centering
 \includegraphics[width=0.449\linewidth]{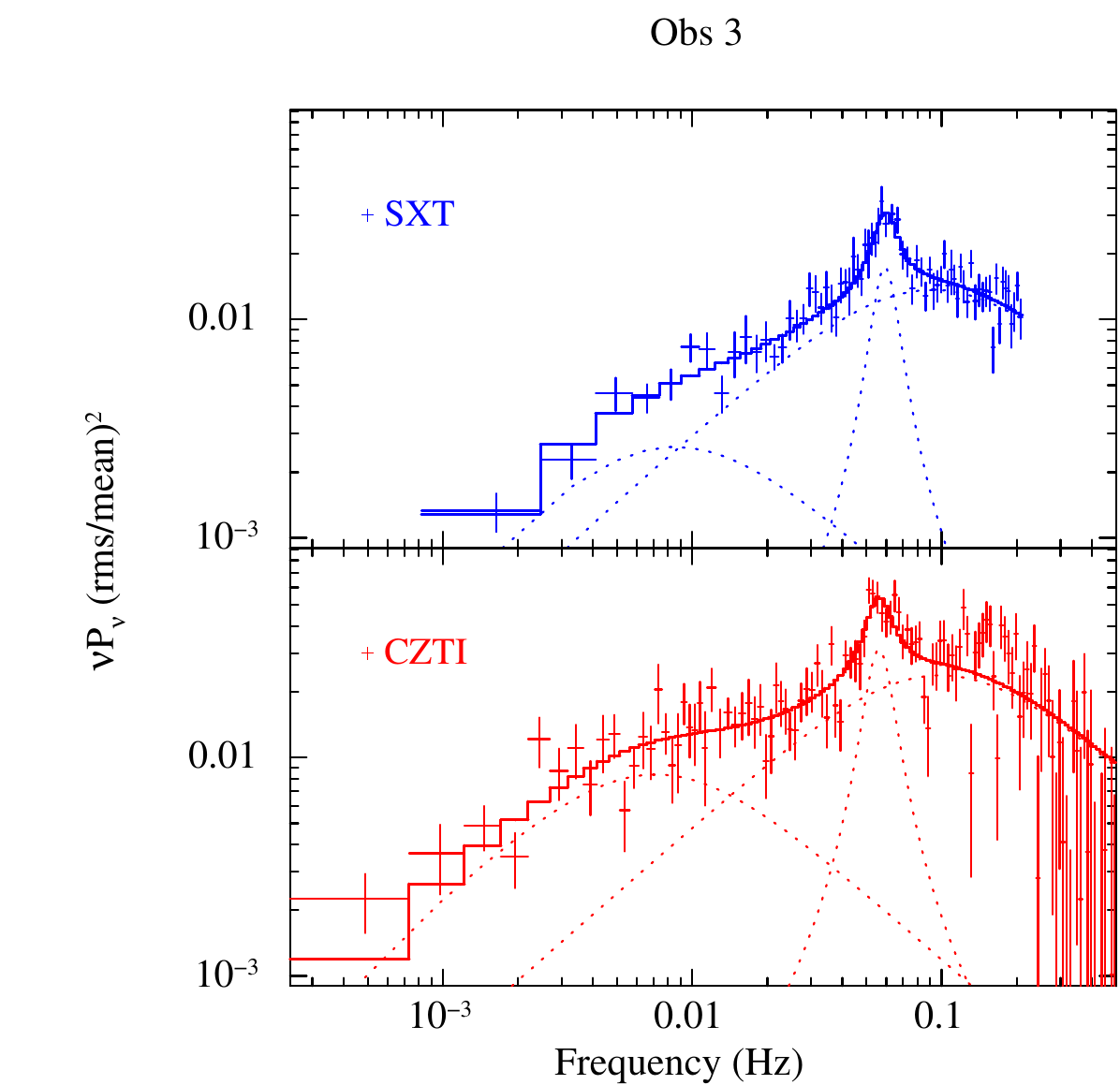}
  \includegraphics[width=0.449\linewidth]{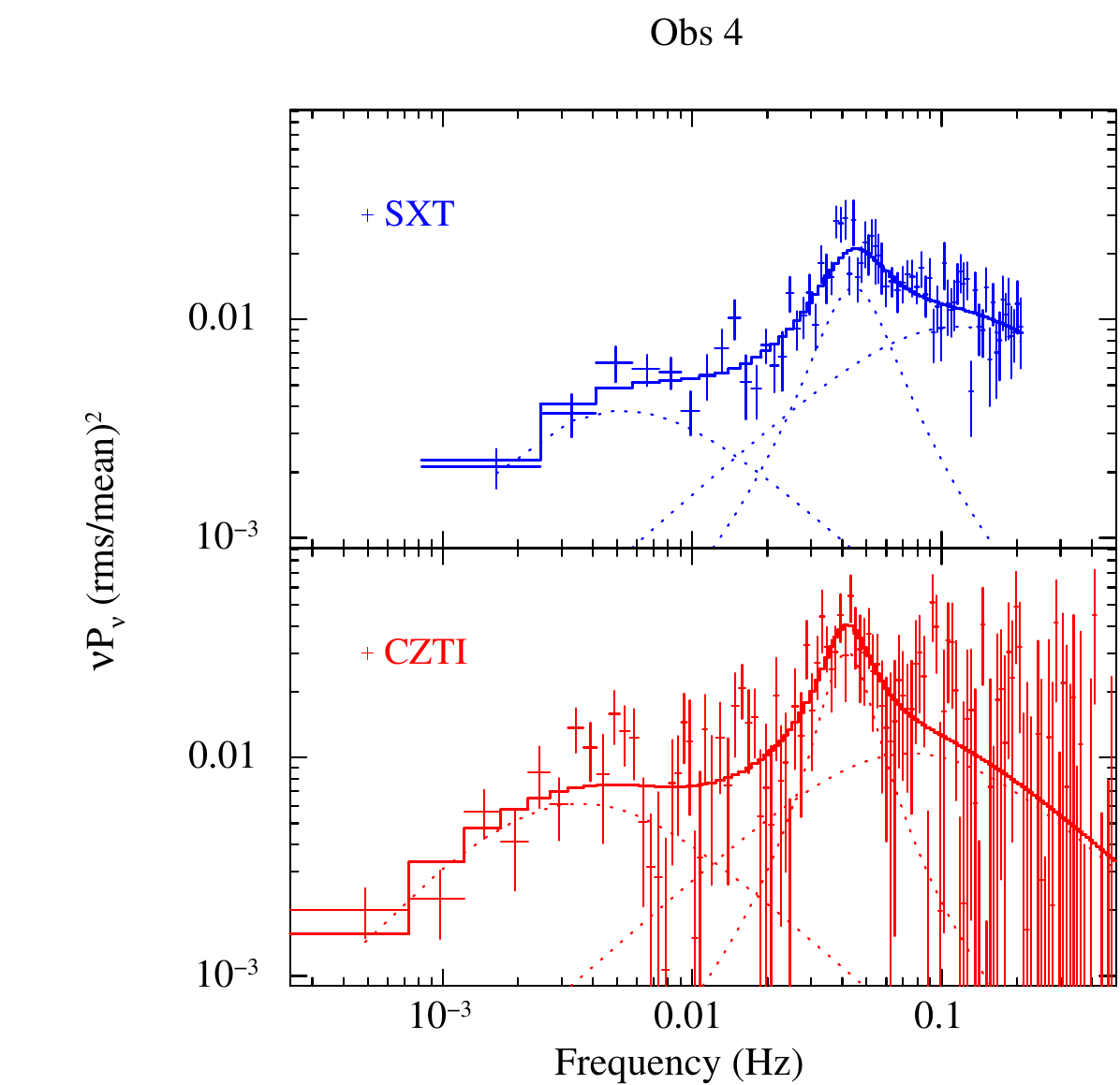}
 \caption{The PDS of \src\ from SXT (in 0.5--7 keV) and CZTI (in 20--150 keV) data of observation 3 (left) and 4 (right). Similar to LAXPC, both PDS can be modelled with three Loretzian functions: two for the noise and one to model the QPO at $\sim$60 mHz and $\sim$40 mHz for observations 3 and 4, respectively.}
\label{fig:qpo-cs}
\end{figure*}

\begin{figure*}
\centering{
\includegraphics[width=\columnwidth]{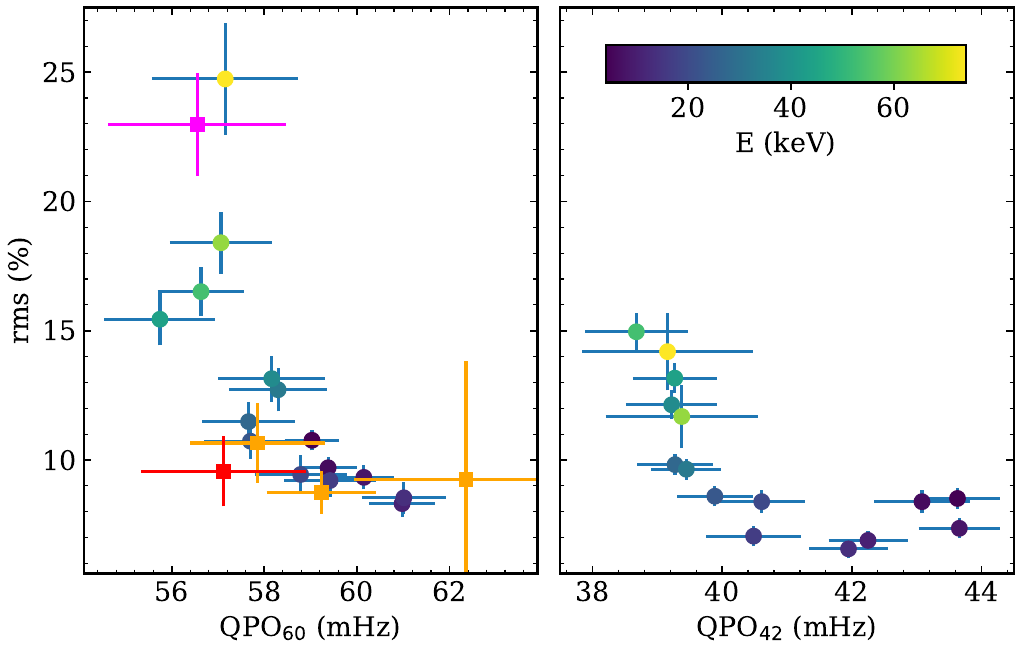}
\caption{The QPO frequency vs. rms amplitude for two different \astrosat-LAXPC observations. The color bar represents the energy. The red and magenta squares are measurements from CZTI for the energy range of 20--50 keV and 50--100 keV, respectively. The orange points represent measurements from SXT for the energy ranges of 0.5--2, 2--4 and 4--7 keV.}
    \label{apx:freq-rms}}
\end{figure*}

%%%%%%%%%%%%%%%%%%%%%%%%%%%%%%%%%%%%%%%%%%%%%%%%%%

% Don't change these lines
\bsp	% typesetting comment
\label{lastpage}
\end{document}